\documentclass[preprint,floatfix,aps,prb]{revtex4-2}

\usepackage[utf8]{inputenc}
\usepackage{amsmath}
\usepackage{amsfonts}
\usepackage{amssymb}
\usepackage{hyperref}
\usepackage[capitalise]{cleveref}
\usepackage{siunitx}
\usepackage{graphicx}
\usepackage{subcaption}
\usepackage{pstricks}
\usepackage{booktabs}

\newcommand{\randspg}{\textsc{RandSPG}\cite{RandSPG}}
\newcommand{\lammps}{\textsc{Lammps}\cite{Lammps}}
\newcommand{\vasp}{\textsc{VASP}\cite{VaspA,VaspB}}
\newcommand{\spglib}{\textsc{SpgLib}\cite{Phonopy}}
\newcommand{\phonopy}{\textsc{Phonopy}\cite{Phonopy}}
\newcommand{\pyiron}{\textsc{pyiron}\cite{Pyiron}}
\newcommand{\mlip}{\textsc{MLIP}\cite{Novikov2021}}
\newcommand{\sphinx}{\textsc{S/PHI/nX}\cite{Sphinx}}

\newcommand{\spg}{\textsc{RandSPG}}
\newcommand{\volmin}{\textsc{VolMin}}
\newcommand{\cellmin}{\textsc{CellMin}}
\newcommand{\intmin}{\textsc{IntMin}}
\newcommand{\hydro}{\textsc{TriAx}}
\newcommand{\shear}{\textsc{Shear}}
\newcommand{\rattle}{\textsc{Rattle}}
\newcommand{\everything}{\textsc{Everything}}

\DeclareSIUnit\angstrom{\text{\AA}}

\begin{document}

\title{
Systematic Atomic Structure Datasets for Machine Learning Potentials:\\Application to Defects in Magnesium
}

\author{Marvin Poul}
\email{poul@mpie.de}

\author{Liam Huber}
\email{huber@mpie.de}

\author{Erik Bitzek}
\email{bitzek@mpie.de}

\author{Jörg Neugebauer}
\email{neugebauer@mpie.de}
\affiliation{Department of Computational Materials Design, Max-Planck-Institut für Eisenforschung GmbH, D-40327, D\"usseldorf, Germany}

\begin{abstract}
    We present a physically motivated strategy for the construction of training sets for transferable machine learning interatomic potentials. 
    It is based on a systematic exploration of all possible space groups in random crystal structures, together with deformations of cell shape, size, and atomic positions.
    The resulting potentials turn out to be unbiased and generically applicable to studies of bulk defects without including any defect structures in the training set or employing any additional Active Learning.
    Using this approach we construct transferable potentials for pure magnesium that reproduce the properties of hexagonal closed packed (hcp) and body centered cubic (bcc) polymorphs very well.
    In the process we investigate how different types of training structures impact the properties and the predictive power of the resulting potential.
\end{abstract}

\maketitle

\section{Introduction}

A key concept in materials science to design materials with tailored properties is defect engineering. In order to successfully employ this concept, one needs a detailed understanding of the relationship between crystal defects on the atomistic scale and their influence on macroscopic materials properties.
Until now this understanding has been provided to a large extent by density functional theory (DFT) calculations especially when investigating e.\,g.\ the thermodynamic stability of materials phases and simple, isolated defects such as vacancies\cite{Tilmann2010}, dislocation arrays\cite{Clouet2009} or high-symmetry planar defects\cite{Finkenstadt2010,Huber2014}.
However, successful defect engineering must include most of the macroscopic and microscopic degrees of freedom of the defects---or risk missing potential candidate states.
Especially in extended defects such as grain boundaries this defect phase space is very large, making it unfeasible to scan with DFT due to its high computational cost and system size restrictions.
Together with recent interest in defect phase diagrams\,\cite{Cantwell2020,KorteKerzel2022} this motivates us to develop a machine learning potential specifically aimed at a transferable description of defects. 
To this end, we will apply the Moment Tensor Potential (MTP) methodology\cite{Novikov2021}, and rigorously examine the impact of training data on the quality and performance of the resulting potentials. 
The approach and the detailed analysis and discussion are however general and can be applied to any machine learning (ML) potential methodology.

Classical potentials are often trained on a set of properties that they ought to reproduce, e.g. relative phase stabilities, surface energies and elastic properties.
The more data hungry machine learning potentials instead use large sets of reference structure with energies, forces and potentially stresses calculated with quantum mechanical models like DFT. 
These reference structures are generally constructed starting from equilibrium structures of interest which are then perturbed in various ways to sample the energy landscape.
This approach can work very well, but can lead to failure of the potential when relevant structures are missing.
Another approach recently presented is to combine active learning and some form for structure generation (randomly, by molecular dynamics or Monte Carlo simulations)\cite{Behler2017,Smith2021,Bernstein2019,Podryabinkin2019}.
By starting from random environments, bias is removed from the training data and then a given active learning algorithm is in control of selecting structures to add to the training set.
For example Smith~\emph{et al.}\,\cite{Smith2021} demonstrates that this works very well for aluminum and it allowed them to obtain a robust potential that predicted the correct relative phase stabilities in a wide temperature and pressure window without any human guidance.
Of these approaches Bernstein~\emph{et al.}\,\cite{Bernstein2019} appears to be most closely related to our approach.
The major difference is the starting point of the generation procedure.
Where they start with completely random distribution of atoms and enforce only a few symmetry operations, we will systematically include most space groups.
Additionally they use an on the fly fitted potential in an iterative scheme to minimize cells whereas we will rely on DFT.

There are also parallels to the work of Podryabinkin~\emph{et al.}\,\cite{Podryabinkin2019}.
The authors' objective there is to predict the stable crystal structures of elements and employ active learning to provide candidate structures that can be investigated with DFT in a reasonable time.
For the explicit purpose of predicting crystal structures they show that this approach works very well.

A challenge in constructing potentials that accurately describe defects is that atoms at or near the defect can have structures that are far away from any low-energy bulk structures. These atoms represent spatially highly localized regions of high energy that are not captured when including only low-energy structure.  
We will discuss in this paper how to construct and utilize  structures that are not energetically near the equilibrium structure.
An approach in that direction is the recent work from M. Karabin~\emph{et al.}\,\cite{Perez2020} and Montes de Oca Zapiain~\emph{et al.}\,\cite{MontesdeOcaZapiain2022}.
Their work aims to sample the descriptor space of the targeted potential model as widely and unbiased as possible.
For this they define a \emph{descriptor entropy} that favors structures with different local environments in the same cell and then maximize or minimize this entropy in a simple annealing procedure.
They show that this yields significantly wider coverage of descriptor space than sets drawn from high temperature MD with a simple size exclusion potential or traditionally constructed training sets, but still includes crystallographic relevant bulk phases and phases.
Since this scheme makes no reference to structure prototypes or crystallography it is a completely unbiased procedure in this sense. 
Instead it relies on the quality of the underlying descriptor set.
This means that changing the descriptor set will produce a different training set.
A potential drawback of their method is also the large number of structures that are generated.\footnote{
    Up to \si{200,000} structures with 32--40 atoms each.
}
In contrast to this we aim here to provide a method for a smaller, descriptor or potential agnostic training set, generated purely based on the constraints placed on atomic positions by the space group symmetries, that still accurately captures the necessary structures.

We structure the paper as follows;
first the construction of the data set in Section~\ref{sec:training}, which is a key concept in this study.
Then a brief review of Moment Tensor Potentials in Section~\ref{sec:mtp}.
In Sections~\ref{sec:cutoff}\,\&\,\ref{sec:fit} we discuss choosing cutoffs and the fitting of the potentials.
Afterwards we verify the potentials on defects and analyse in detail the influence of training data in Sections~\ref{sec:verification}\,\&\,\ref{sec:stability_discusion}.
This analysis demonstrates the performance of our main idea.
We close with a brief comparison with active learning in Section~\ref{sec:al} before we summarize our findings in Section~\ref{sec:conclusion}.

\section{Methods}\label{sec:method}

\subsection{Training Set Construction}\label{sec:training}

\begin{figure*}
    \includegraphics[width=\textwidth]{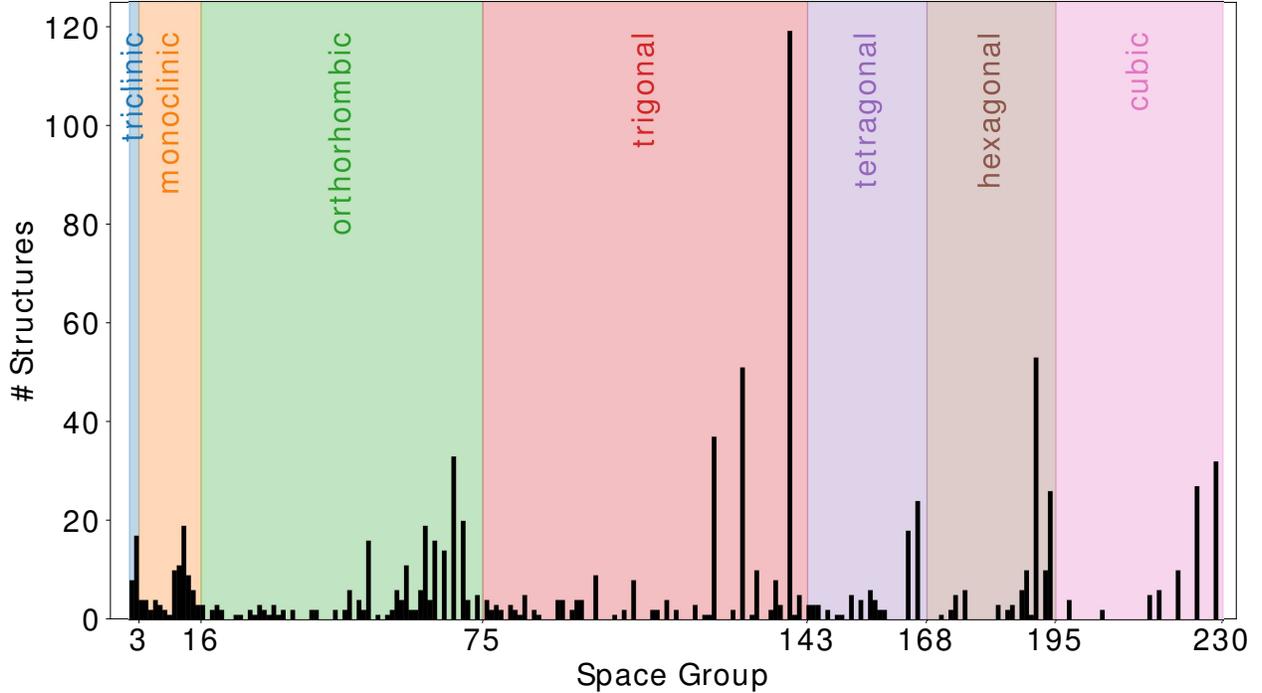}
    \caption{Frequency of crystal systems in the \spg set (see text).}
    \label{fig:spghist}
\end{figure*}

We construct several different training subsets, each of which explores slightly different regions of phase space that have clear physical interpretation.
We generate these sets in an iterative, multistage process.
The foundational dataset for this process, which feeds into all further subsets, consists of random (periodic) crystal structures obtained from \randspg.  
We generate these structures with one to ten magnesium atoms per cell, asking for all possible space groups, 1-230, and allowing volumes $\pm10\,\%$ around the equilibrium atomic volume of hcp Mg of $\Omega_0 = \SI{22.87}{\angstrom^3/atom}$, obtained from DFT at $T = \SI{0}{K}$.
We label this the \spg\ set. 
Note that this approach requires as only input parameters the volume range and number of atoms considered in super cells for training.
An automation and extension to other materials is therefore straightforward.
Applying this approach, not all space groups are present, because some symmetries are not consistent with the allowed volume range or lead to structures with very inhomogeneous particle distribution.
As a check the space groups have been determined with \spglib.
\Cref{fig:spghist} shows the frequency of each space group and crystal system in our initial data set.
While not all systems are equally present, there is a sizeable number of structures available for each.

From this starting point, we then successively minimized the volume, cell shape, and the internal coordinates independently using \vasp.  
These calculations are done at low convergence parameters, since they only serve to bring the structures near the equilibrium structures and the energies from these runs do not enter the fitting routines.
We call these sets \volmin, \cellmin\ and \intmin\ respectively.

Naturally the minimization generally leads to higher symmetry structures exploring a reduced phase space.
In fact, in the fully internally relaxed set some space groups are no longer present.
This reduces the inherent dimensionality of the minimized training sets, but we have not attempted to filter structures that relaxed into the same minima.
It is also noteworthy that volume minimization---particularly for the structures with more atoms per unit cell---can lead to quasi 1D and 2D structures.
This gives the potentials the opportunity to see structures resembling surfaces and isolated atoms even though in the construction setup we do not explicitly enforce such structures.\footnote{
    For materials with dynamically unstable ground states or phases that are stable only at elevated temperatures, this minimization procedure might appear to under sample or neglect them.
    However even in these cases as long as the initial \spg{} set is sufficiently diverse to find the structures once, their local environment will be sampled by the minimizing training set.
}

As a final step in our process for creating training data, the structures from \intmin\ are disturbed by either a random triaxial (\hydro, up to 80\,\%) strain, a combination of random shear strains (\shear, up to 80\,\%), or by random displacements of atoms combined with a small random strain tensor (\rattle, \SI{0.5}{\angstrom} mean displacement and up to 5\,\% strain).\footnote{
    For the \hydro~and \shear~sets the respective entries in the strain tensor are chosen from uniform random distribution in range $\pm80\,\%$.
}
For each structure in \intmin~these modifications are applied five times, resulting in five times more structures for the respective derived training sets, \hydro, \shear~and \rattle~than in the \intmin~set.

\begin{figure}
    \scriptsize
    \centering
    \def\svgwidth{\linewidth}
    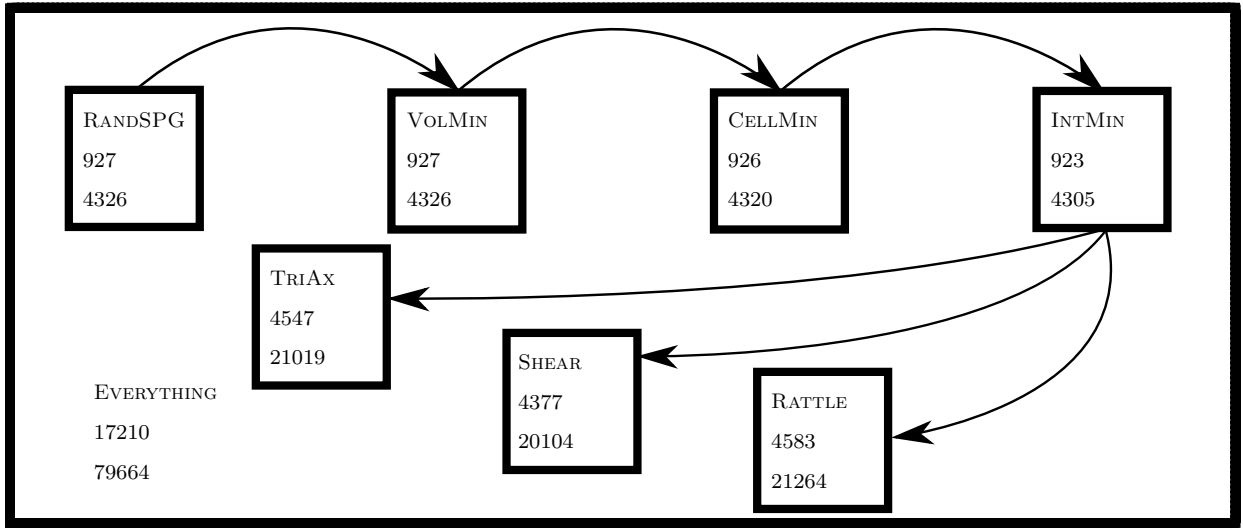
    \caption{
        Schematic procedure to generate the training sets.
        First number gives the number of structures in each set, the second the total number of atoms.
    }
    \label{fig:training_schematic}
\end{figure}

\Cref{fig:training_schematic} gives a conceptual overview of this procedure.  
During each step some structures resist DFT calculations due to excessively close atoms or deformed cells, which we then discard.
Also shown in \cref{fig:training_schematic} are the number of structures (top number) and atoms (i.\,e.\ atomic environments, bottom number) in each structure set.
We will examine the performance of potentials fitted to each of these sets compared to potentials fitted to the set of all structures, \everything.

\subsubsection{Training Sets for Multi Component Potentials}

The procedure can be extended to binary or ternary compounds by using \randspg~to generate structures with various concentrations.
This would naturally increase the number of structures substantially.
It is not clear at this point how dense in concentration space such training sets would need to be or how well potentials would be able to interpolate or extrapolate between concentrations.
In this case it might be necessary to combine the data generation procedure shown here with data selection strategies from active learning schemes.
However, for now we focus on unaries and leave the exploration of alloys to future work.

\subsection{Test Data}

In general in machine learning it is custom to reserve some percentage of the reference data in a hold-out set on which to test the final model.
We explicitly decide against doing this and will use the full data set for fitting and report only the errors on this (training errors).
The reason is that to test the potentials we construct completely fresh data closer to the application domain of the potential, such as defect structures, phonon and elastically strained calculations.
The results of this testing are discussed in Section~\ref{sec:verification}.
We chose to do this because the meaningfulness of test errors depends to a large extent on the sampling of reference structures.
If the reference structures are not drawn evenly from the full space that the potential will be applied on, the train-test errors can give the impression that the potential is fitted well, even though there are gaps in the potential, simply because the relevant structures never entered the reference structure set.
This is discussed in more detail, e.\,g.\ in a review from J.\,Behler~\cite[Fig.\ 10 and the discussion Sec.\ 4]{Behler2017}.
Since we now rely on these separate completely out-of-fold structures for testing we opt to include the full reference data set in the fitting to provide more learning opportunities to the potential.

\subsection{DFT Data Generation}

All training data is generated using \vasp{} using the Projector Augmented Wave (PAW) method\,\cite{Blochl1994,Kresse1999} and the PBE\,\cite{PBE} functional with the standard s-valent pseudopotential from the \vasp distribution.
$\Gamma$ centered $k$-meshes with $27\times27\times27$ $k$-points and plane wave energy cutoff of \SI{550}{eV} are used.
While the structures vary in volume, we keep the $k$ points constant to avoid discontinuities in the potential energy surface.
The chosen $k$ point setting corresponds to a $k$ mesh spacing of \SI{0.06}{\angstrom^{-1}}.\footnote{
    Less then 10\,\% of the structures have $k$ mesh spacing larger then \SI{0.09}{\angstrom} and very few less than the above quoted.
}
All calculations used the Methfessel-Paxton occupation smearing scheme of order 1 with a smearing parameter of \SI{0.2}{eV}.\,\cite{Methfessel}
By convergence testing we find the energies to be converged to \SI{0.6}{meV} and the forces to \SI{7e-5}{eV/\angstrom}.
These values represent the mean error of the training calculations with respect to a sample of 50 structures of each training set (350 in total) calculated at a $37\times37\times37$ k-point mesh and a plane wave cutoff of \SI{687.5}{eV}.
DFT data for the verification calculations is generated with the same parameters except for large grain boundary and surface structures where we use a $k$-mesh spacing of \SI{0.05}{\angstrom^{-1}}.

A small number of calculations fail during the minimizations and the final training set generation.
They are automatically discarded and do not enter subsequent steps.
Figure~\ref{fig:training_schematic} shows that their total number is small however.

\subsection{Moment Tensor Potentials}\label{sec:mtp}

Moment Tensor Potentials (MTP) are machine learning interatomic potentials originally introduced by A.~Shapeev.\cite{Shapeev2016}
We will briefly review this formalism here, but leave the details to to the original authors.\cite{Novikov2021}

The total energy, $E^\text{MTP}$, of any atomic structure is constructed from contributions of neighbors around each atom, which are expanded in linear basis functions
\begin{equation}
    E^\mathrm{MTP} 
    = \sum_i^N V(\mathfrak{n}_i)
    = \sum_i^N \sum_\alpha \xi_\alpha B_\alpha(\mathfrak{n}_i),
\end{equation}
where $\mathfrak{n}_i$ is the atomic environment around atom $i$, $N$ is the total number atoms, $B_\alpha$ are the descriptor basis functions and $\xi_\alpha$ the linear expansion coefficients, which are determined during the fitting procedure, and $\alpha$ runs over all basis functions.\footnote{
    Readers acquainted more with classical interatomic potentials might expect a factor $\frac{1}{2}$ in front of the per atomic contribution to the total energy, to avoid double counting.
    In the usually machine learning formalism, this factor is absorbed into the basis coefficients $\xi_\alpha$.
}
The descriptor basis functions are defined as contractions of the \emph{Moment Tensors}, $M_{\mu, \nu}$, defined in the single-component case as
\begin{equation}
    M_{\mu,\nu}(\mathfrak{n}_i)
    = \sum_{j \in \mathfrak{n}_i} f_\mu(\left|r_{ij}\right|) \overbrace{r_{ij}\otimes\dots\otimes r_{ij}}^{\nu~\text{times}}
\end{equation}
where $r_{ij}$ is the vector connecting the $i$'th and $j$'th atoms and $\otimes$ is the outer product on vectors and tensors.
The radial functions $f_\mu$ are expanded in an orthogonal polynomial basis and contain an outer cutoff $R_c$, such that their derivatives go smoothly to zero.
This encodes the locality assumption generally made in interatomic potentials.
The polynomial expansion coefficients of the radial functions are also additional fitting parameters.
The authors then \emph{define} the \emph{level} of an MTP as
\begin{equation}
    \mathrm{lev}\,M_{\mu,\nu} = 2 + 4\mu + \nu \quad.
\end{equation}
The level of the basis functions $B_\alpha$ are then the sum of the levels of the tensors out of which they are contracted.
Finally the potentials are constructed by including all basis functions below a given level $l_\mathrm{max}$.
This implicitly defines up to what values of $\mu$ and $\nu$ the Moment Tensors $M_{\mu,\nu}$ are included in the final potential.
The number of fitting parameters in a potential goes exponentially with its level.

\subsection{Cutoff Radius Determination}\label{sec:cutoff}

\begin{figure}
    \centering
    \includegraphics[width=\textwidth]{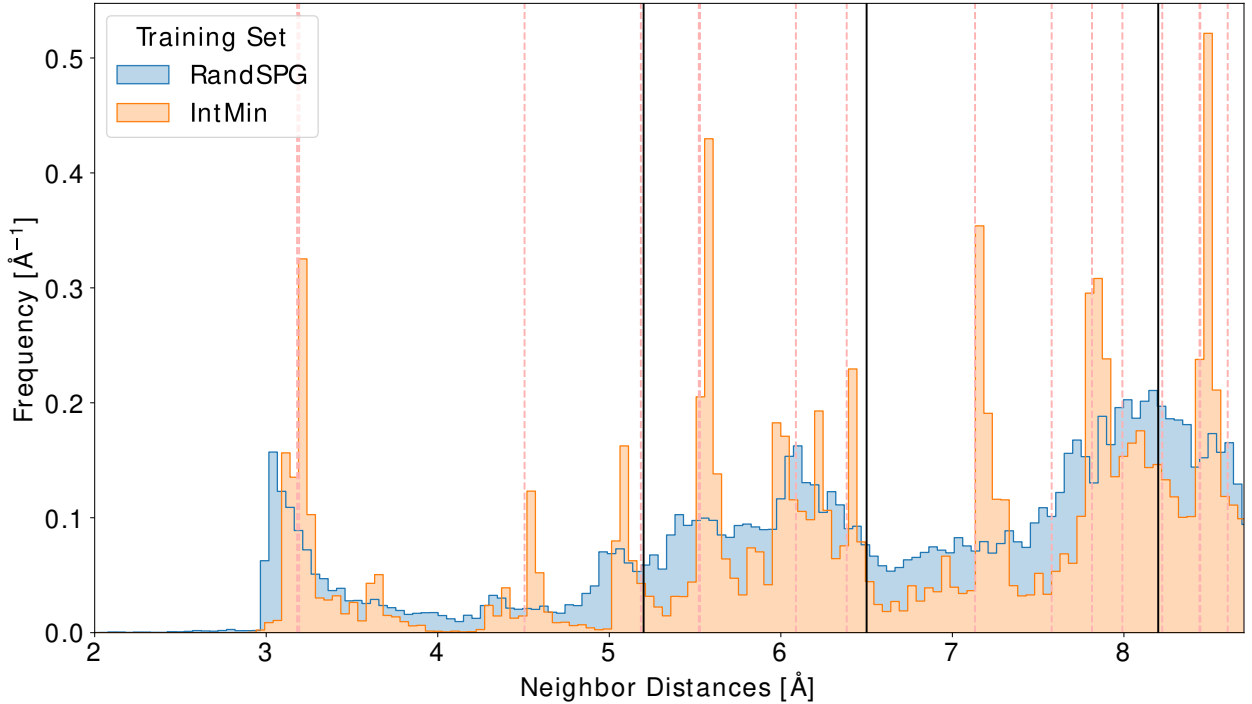}
    \caption{
    Histogram of neighbor distances in \spg{} and \intmin{} training sets.
    Black lines are the considered cutoffs; red dashed lines the shells of hcp Mg at equilibrium volume of $\Omega_0$.
    }
    \label{fig:cutoffs_comparison}
\end{figure}

MTPs are local potentials, i.\,e.\ they separate the total energy of a structure into individual contributions of each atom or, more specifically, to spatially localized environments that are atom-centered.
This environment is defined by a lower and upper cutoff radius, $R_c$, such that all the individual regions of space considered are shell-shaped.  
The first task in fitting potentials then is to determine appropriate cutoffs. To this end, we first calculate nearest neighbor vectors for all structures in the training sets.
This task requires a detailed and explicit analysis. 
As the lower cutoff we pick \SI{1.8}{\angstrom} for all potentials which is the pseudo-potential cutoff used in our \vasp{} calculations.
We select a set of upper cutoffs which we thoroughly investigate to determine their impact on the potential's accuracy.
\Cref{fig:cutoffs_comparison} shows the distribution of neighbor distances in the \spg{} and \intmin{} training sets.
Also drawn are the hcp Mg shell distances at $\Omega_0$ (dashed red lines) and the three considered cutoffs (black solid lines).  
Between \SIrange{2}{3}{\angstrom} only very few structures are present due to the constraints we have put on the structure generation.
While the nearest neighbor distances of the \spg{} structures are mostly evenly distributed, the \intmin{} distributions shows distinct peaks.
This is expected after energy minimization and gives important clues what cutoffs are physically meaningful.
The peaks tend to align with the hcp shell distances (red dashed lines), but additional peaks from other structures are also present.
It can be seen that $R_c = \SI{5.2}{\angstrom}$ includes the first three shells, $R_c = \SI{6.5}{\angstrom}$ the first six, and $\SI{8.2}{\angstrom}$ the first ten shells. 
We will pick these cutoffs for the rest of the paper.
The choice of the cutoff has important consequences on the quality of the potentials as will be seen in Section~\ref{sec:volume} and it is therefore important to explicitly check what cutoffs may reasonably be considered without depriving the model of physically relevant information.
Finally the fact that \spg{} has a fairly smeared out distribution is also important as it gives the potentials critical information on out-of-equilibrium configurations.

\subsection{Fitting Procedure}

We fit MTP models for each of the data sets at different model complexities, choosing levels from 8 to 24 with the \mlip~program, which performs energy, force, and stress matching in a least-squares optimization.
All potentials are fitted with respect to energies, forces and stresses from DFT, with weights of $1$, $0.01$ and $0.001$ respectively.

\section{Results and Discussion}

\subsection{Fitting Results}\label{sec:fit}

\begin{figure}
    \centering
    \begin{subfigure}{\textwidth}
        \includegraphics[width=\textwidth]{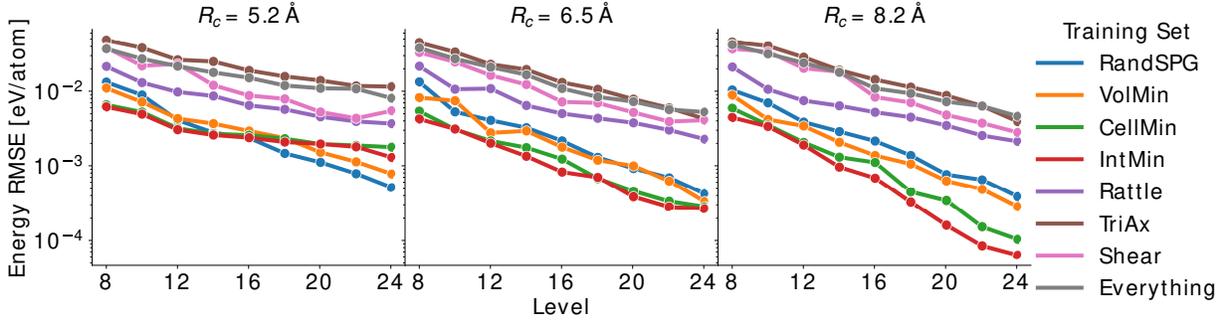}
        \caption{
            Energy RMSE over the level of the potentials. 
        }
        \label{fig:energy_errors_sub_levels}
    \end{subfigure}
    \begin{subfigure}{\textwidth}
        \includegraphics[width=\textwidth]{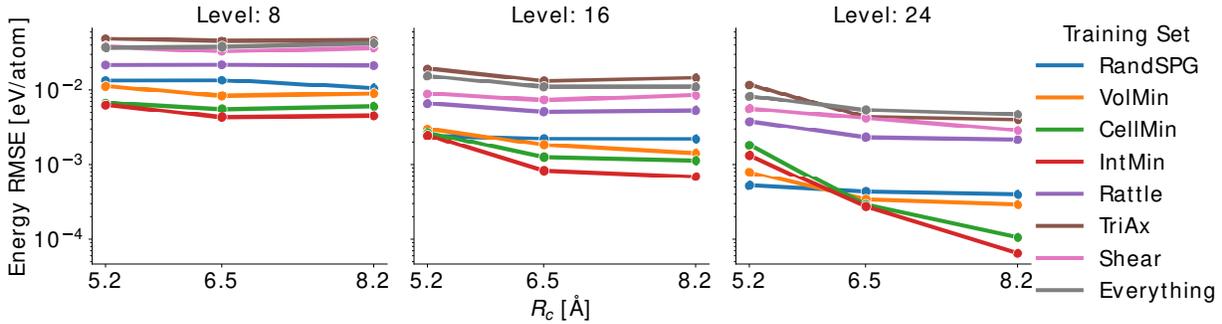}
        \caption{
            Energy RMSE over the cutoff radius of the potentials.
        }
        \label{fig:energy_errors_sub_rmax}
    \end{subfigure}
    \caption{
        Energy training RMSE of the potentials fitted to different training sets over potential level and cutoff.
        Subplots show potential level and cutoff radii.  
        Higher cutoffs clearly show improved convergence rates at high level.
    }
    \label{fig:energy_errors}
\end{figure}

We obtain energy, force and stress root mean square errors (RMSE) values after each fit.
Energy RMSE are plotted in \cref{fig:energy_errors_sub_levels} as a function of potential level for the three cutoffs.
They follow a systematic improvement, but interestingly the different structure sets follow a different convergence.
Since the training sets contain progressively minimized structures their structural complexity decreases and they appear to become easier to capture for the potentials.
The trend only reverses with the strained and displaced sets which add complexity again.
The lowest RMSE at the highest level also follow this trend, from which we conclude that potentials fitted to larger, more diverse, structure sets naturally have a higher interpolation error than potentials with smaller training sets.

In \cref{fig:energy_errors_sub_rmax} the same energy RMSE is plotted as a function of cutoff for three selected levels. 
It can be seen that the levels below 24 quickly saturate with respect to the cutoff, i.\,e.\ to the low level descriptors higher cutoffs do not necessarily include more information. 
Since the level characterizes both the body-order and the number of radial basis functions included in the potential, it is not clear which of them is the limiting factor.\footnote{
    The \textsc{Mlip} code allows to train potentials where body order and number of radial functions are independently varied, but we have not investigated in this work.
}
Thus, to optimize the numerical performance of a potential one should carefully check whether for the given descriptor level the cutoff is appropriately chosen.\footnote{
    Comparisons of the runtime cost as a function of cutoff radius are explicitly given in \cref{fig:runtime}.
}

\subsection{Error-Cost Tradeoff}

\begin{figure}
    \centering
    \includegraphics[width=.8\textwidth]{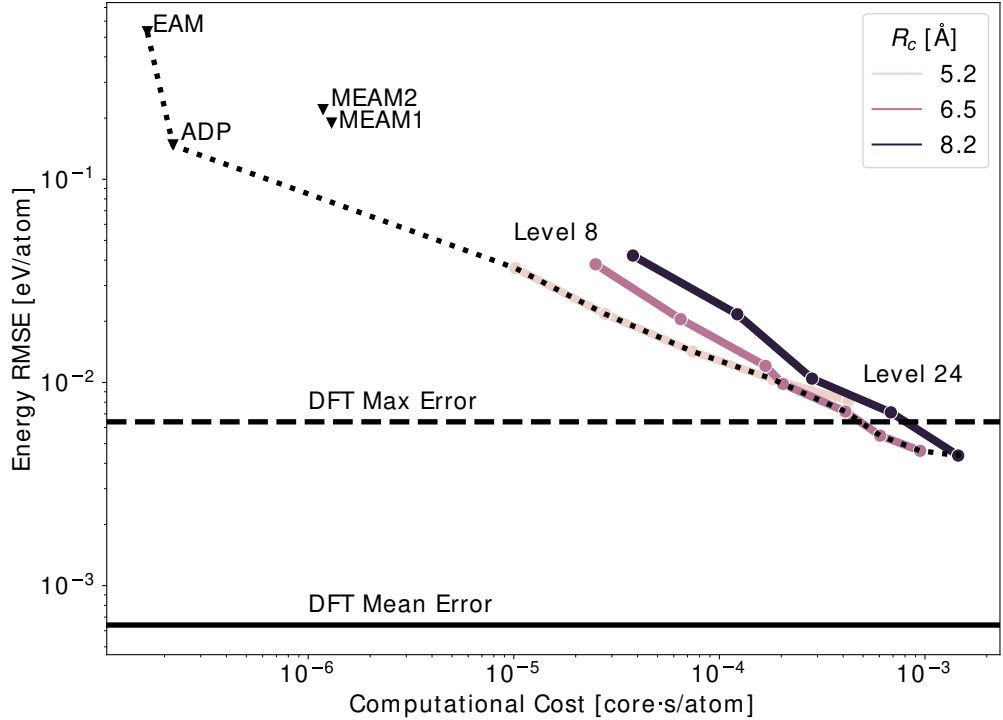}
    \caption{
        Time per force call per atom versus root mean square error of the energy on the \everything~set.
        Colors symbolize the cutoff radius.
        Each line is constructed by using potentials of increasing level, from 8 to 24. 
        The dotted line marks the Pareto front.
        The horizontal black lines indicate the mean (solid) and maximum (dashed) errors in the DFT training set from convergence testing.
        The classical potentials (ADP\,\cite{Smirnova2018}, EAM\,\cite{Sun2006}, MEAM1\,\cite{Dickel2018}, MEAM2\,\cite{Kim2009}) for comparison are shown with black triangles.
    }
    \label{fig:pareto}
\end{figure}

Once sets of potentials are obtained, practitioners must pick which to use for certain applications.
One way of making that choice is to look at the trade-off between computational cost and their accuracy.
Here we use the training RSME as measure for the accuracy of the potentials fit to the \everything~set.
To provide a measure for the computational cost we run NVT-MD on $6\times6\times6$ primitive hcp Mg unit cells for \si{10000} steps, or \SI{10}{ps}, at \SI{50}{K}.
Additionally we calculated the RMSE for four classical potentials\cite{Sun2006,Dickel2018,Smirnova2018,Kim2009} on the \everything{} set and their runtime in the same MD setup.
\Cref{fig:pareto} shows the Pareto front for these potentials where we compare the runtime to the fitting error.
Drawn as horizontal lines are the aforementioned DFT convergence errors; once the mean error (\SI{0.6}{meV/atom}, solid line) and the maximum error (\SI{6.4}{meV/atom}, dashed line).
Generally the MTPs are 1-2 orders of magnitude more accurate, while being 1-3 orders slower than the classical potentials.
None of the classical potential achieve lower errors than \SI{100}{meV}.
At low-cost and low-accuracy, the Pareto front follows the potential fit with a cutoff $R_c = \SI{5.2}{\angstrom}$ before switching over to $R_c = \SI{6.5}{\angstrom}$ at around \SI{10}{meV/atom} error.
While the $R_c = \SI{8.2}{\angstrom}$ potential is behind the Pareto front (with the exception of the highest level, where it is marginally more accurate than $R_c=\SI{6.5}{\angstrom}$, we will later see in Sections~\ref{sec:gb} and \ref{sec:volume} that it is still useful due to superior performance when treating defects and different closed packed structures.
The plain computational cost of each potential as a function of level and cutoff are shown in appendix~\ref{sec:appendix-pareto} for all potentials fit to \everything{}.

\subsection{Verification}\label{sec:verification}

As mentioned in the methods, we do not split the fitting data into traditional train and test sets. 
Instead we perform calculations of various quantities that we can compare to independent (i.\,e.\ not entering the fitting) DFT calculations.
In this section we will focus on results for the potentials fitted to \everything{} unless otherwise noted, deferring the discussion of the performance of the various data sets to Section~\ref{sec:stability_discusion}.
In total more than 1000 additional structures have been calculated with DFT for the verification calculations below.
These structures are part of volumetric and uniaxial strain, phonon and defect calculations are explained in more detail below.

\subsubsection{Strain Calculations}\label{sec:volume}

\begin{figure}
\includegraphics[width=\textwidth]{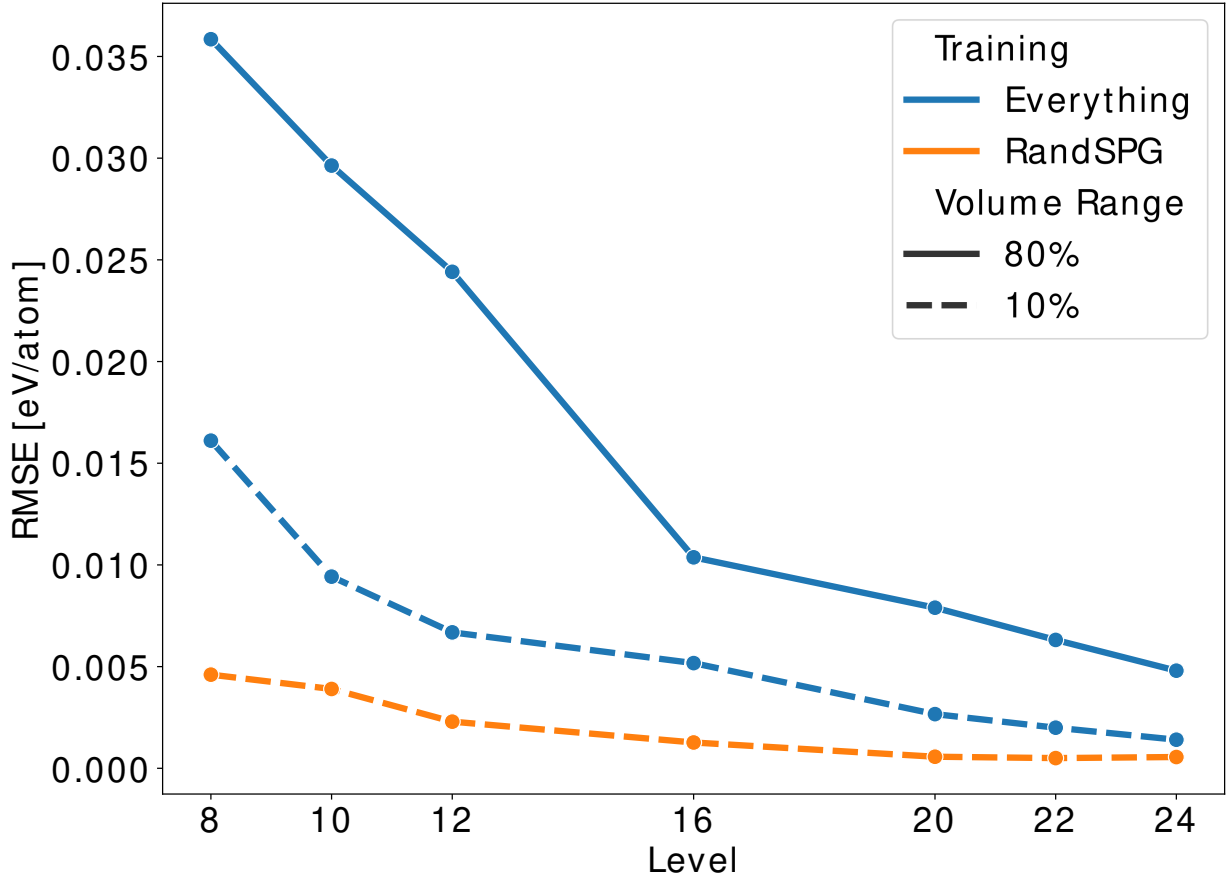}
\caption{
    Energy RMSE of potentials fitted to \spg~and \everything~ with $R_c=\SI{8.2}{\angstrom}$ averaged over the volume ranges (10\,\% and 80\,\% respectively) and hcp, fcc, dhcp and bcc as a function of potential level.
    While the simpler \spg{} set achieves lower errors on the narrower range, it completely fails at the larger range, where as potentials trained on \everything{} still achieve less than \SI{10}{meV} over the wide range.
    The RMSE on the 80\,\% volume range for potentials fitted to \spg~is not shown as it exceeds \SI{1}{eV/atom} indicating clear failure of these potentials on larger volume ranges.
    Potentials with $R_c = \SI{5.2}{\angstrom}$ and $R_c = \SI{6.5}{\angstrom}$ show the same qualitative behavior.
}
\label{fig:vol_range_rmse_comparison}
\end{figure}

An important part of verifying machine learning potentials is checking that the stability of the bulk phases is correctly predicted over the volume range of interest.
To this end we calculate the E-V curves of hcp-, fcc-, dhcp- and bcc-Mg.
First we strain the reference structures hydrostatically within $\pm80\,\%$ and $\pm10\,\%$ of the hcp equilibrium volume.  
\Cref{fig:vol_range_rmse_comparison} shows the RMSE on these ranges as a function of potential level and cutoff.
We compare here potentials fitted to \spg{} and \everything{}.
Not shown is the error of the \spg{} set on the 80\,\% range because this potential clearly failed, as the error exceeds \SI{1}{eV/atom}.
The \everything{} set achieves \SI{\approx 5}{meV/atom} in the 10\,\% range and \SI{< 10}{meV/atom} in 80\,\% range.

\begin{figure} 
\includegraphics[width=\textwidth]{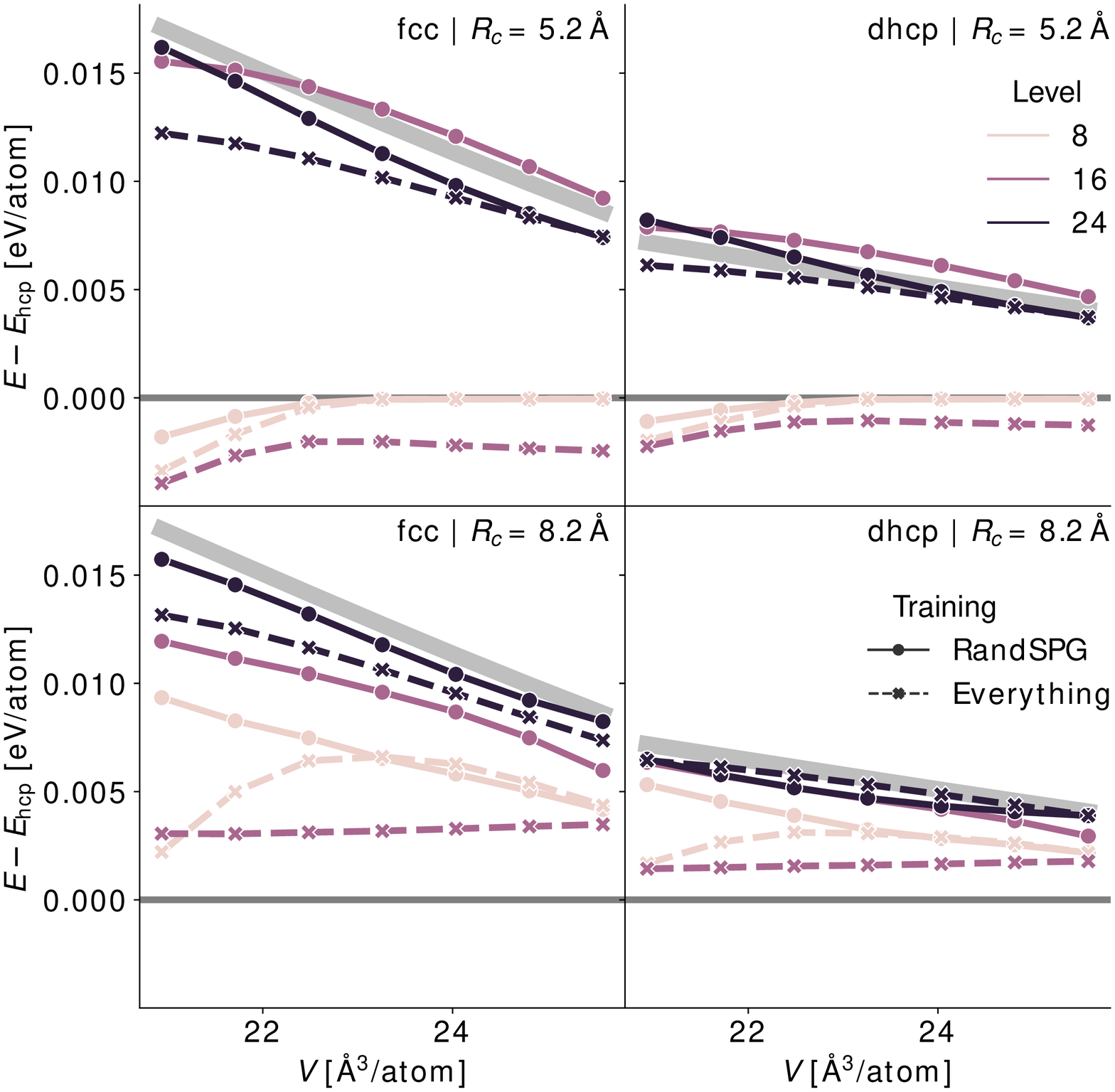}
\caption{
    Energy difference to HCP per atom vs atomic volume for FCC and DHCP (columns); 
    think gray lines are DFT reference energies;
    figures in the top row correspond to potentials with $R_c = \SI{5.2}{\angstrom}$ and in the bottom row $R_c = \SI{8.2}{\angstrom}$.
    The potentials are fit to the \spg{} (solid) and \everything{} (dashed).
    While both sets and cutoffs eventually converge to the DFT phase stabilities, low cutoffs and low ranks clearly fail in differentiating close packed structures.
}
\label{fig:mtp_stacking_failure}
\end{figure}

\Cref{fig:mtp_stacking_failure} shows the energy of dhcp and fcc relative to the predicted hcp energy for potentials of level 8, 16 and 24 with cutoff \SI{5.2}{\angstrom} and \SI{8.2}{\angstrom}, and training sets \spg{} and \everything{}.
All potentials eventually converge close to the DFT values, but note the pronounced failure of level $8$ potentials at the lower cutoff.
Even more interestingly, for the larger training set also the level 16 potential fails to distinguish the three structures.
At higher cutoffs all levels are able again to differentiate the structures, though again the larger training set has a harder time correctly describing all structures.
The first observation implies that interpolation errors become relevant.
Since the \everything{} set is much broader in phase space, we interpret the failure of the level 16 potential as still having too few basis functions to span the large configurations space covered by \everything{}.
We will return to this in the context of active learning in Section~\ref{sec:al}.
This additional interpolation error for larger training sets comes at the advantage of a larger applicability as can be seen referring back to \cref{fig:vol_range_rmse_comparison}.
While the smaller \spg{} set outperforms \everything{} in the smaller volume range, it is not at all usable on the larger range. 

The bottom row in \cref{fig:mtp_stacking_failure} shows the value of larger cutoffs, even though \cref{fig:pareto} did not seem to indicate that earlier.
Here potentials with larger cutoff do no fail at differentiating the close packed structures (though quantitative agreement is achieved only at higher levels).
The larger cutoff appears to make lower basis functions more efficient at differentiating closed packed structures.
This also helps in the computational efficiency in a round-about way, as e.\,g.\ a level 12 potential with cutoff \SI{8.2}{\angstrom} takes as much time per force call as a level 20 potential with cutoff \SI{5.2}{\angstrom} while being less prone to overfitting, see \cref{fig:runtime}. Thus, to construct potentials with an optimal balance between computational efficiency and accuracy the two parameters---levels and cutoff---should be simultaneously optimized.

Additionally we strained the prototype structures along each of the six possible axes (three strain, three shear) also within $\pm60\,\%$ and compared again DFT and MTP.
For space reasons we do not show the results here and defer to Section~\ref{sec:stability_discusion}

\subsubsection{Phonons \& Force Constants}

After checking static properties we now turn to dynamical properties.
We have calculated phonon spectra and band structures for hcp and bcc cells at the minimum energy volume, as well as bcc cells compressed to \SI{12}{\angstrom^3/atom} where it is dynamically stable.
All calculations were performed with \lammps{} and \phonopy{} using an interaction cutoff of \SI{10}{\angstrom}, which corresponds to a $4\times4\times4$ supercell for hcp and a $5\times5\times5$ supercell for bcc.

\begin{figure}
    \centering
    \includegraphics[width=\textwidth]{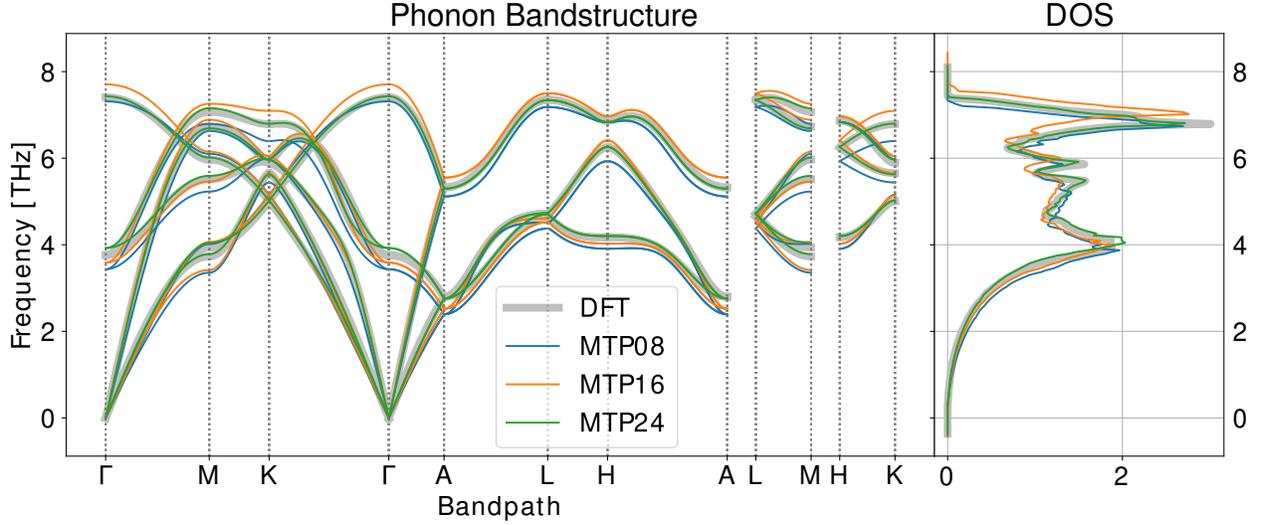}
    \caption{
        Phonon band structure and density of states calculated with DFT (black line and dots) and three MTPs at levels 8, 16 and 24 (colored lines).
        The MTPs show very good agreement with DFT, increasing in accuracy with level.
    }
    \label{fig:hcp_phonons}
\end{figure}

\Cref{fig:hcp_phonons} shows the phonon band structure and density of states calculated with DFT and three MTPs fitted on \everything{} with cutoff \SI{8.2}{\angstrom} at three levels: 8, 16 and 24.
The potentials fitted on \everything{} show very good agreement with the DFT results.
Our validation results also indicate a significantly better description of bcc Mg, both in the compressed high pressure state as well as at the equilibrium volume, as compared to other recently reviewed Mg potentials.
The band structure and density of states for bcc Mg are shown in Appendix \ref{sec:app_phonon}.
Troncoso \textit{et al.}~\cite{Troncoso2022} review this topic and find MEAM potentials are the best so far to study the dynamical behaviour of bcc Mg, but also report that the same potentials are deficient in their elastic properties.\footnote{
    While we have not shown the elastic constants calculated from our potential here, it is implicit in the very good agreement of the energies for large volume strains shown in Section~\ref{sec:volume}.
}
Neural Network Potentials\,\cite{Stricker2020,Behler2007} reviewed in the same paper are found to be better for this application, but still predict wrong dynamical instabilities or predict them in the wrong part of the band structure. Based on these findings the authors concluded that these properties could be improved by specifically targeting bcc structures in the training set.\,\cite{Troncoso2022}
The fact that our potentials reproduce the dynamical as well as elastic properties so well without specifically targeting these values during the training set generation nicely illustrates the ability of our training approach to produce transferable potentials.

\begin{figure}
    \centering
    \includegraphics[width=\textwidth]{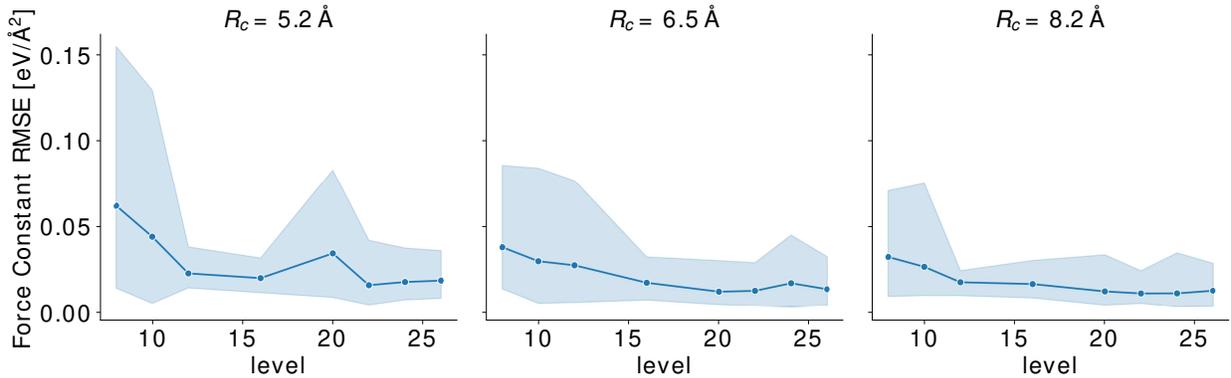}
    \caption{
        Error in the force constants of potentials fit to \everything{} for different cutoffs.
        Lines are the averaged RMSE over all structure protoypes and
        shaded areas indicate their spread.
        Higher levels clearly improve accuracy, while higher cutoffs lead to a more consistent description, i.\,e.\ lower spread between structures.
        At highest levels the potentials reach force constant accuracy between \SIrange{0.01}{0.05}{eV/\angstrom^2}.
    }
    \label{fig:force_constants_rmse}
\end{figure}

For a quantitative comparison between DFT and MTP phonon spectra we use the RMSE in the force constants computed on the same supercell.
We refrain from directly comparing the resulting density of states, since such a comparison is ambiguous when the domain of the compared densities does not match.
In contrast, errors in the force constants are naturally related to the errors in the forces themselves and are therefore an interesting quantity for validation.
\Cref{fig:force_constants_rmse} shows the force error as a function of potential level averaged over the structure prototypes.
Shaded areas indicate the spread of the errors over the prototypes.
For all cutoffs the accuracy naturally increases with level.
We also see, however, that larger cutoffs aid the potentials in making more consistent predictions, i.\,e.\ yield lower spread of the errors over the structure prototypes.
Assuming an average thermal displacement of atoms in the tenths of $\text{\AA}$, these quantities mean that the average thermal force will have errors in around \SI{0.1}{eV/\angstrom}, which is larger than the DFT convergence error, but similar to the training force RMSE values.

\subsubsection{Vacancies}

The vacancies are constructed using $3\times3\times3$ super cells and the reference bulk energies are calculated on the corresponding defect free primitive cells. 
We do this for each of the four structure prototypes mentioned previously: hcp, fcc, dhcp and bcc.  
DFT calculations are run with a plane-wave cutoff of \SI{550}{eV} and k-mesh spacing of \SI{0.05}{\angstrom^{-1}}.\footnote{
    This corresponds to Monkhorst-Pack meshes of $15\times15\times8$ for hcp defect supercell, $15\times15\times4$ for dhcp and $16\times16\times16$ for fcc and bcc.
}
Structures are relaxed in their internal degrees of freedom and volume in DFT and MTP separately before the formation energy is calculated.

\begin{figure}
    \centering
    \includegraphics[height=.6\textheight]{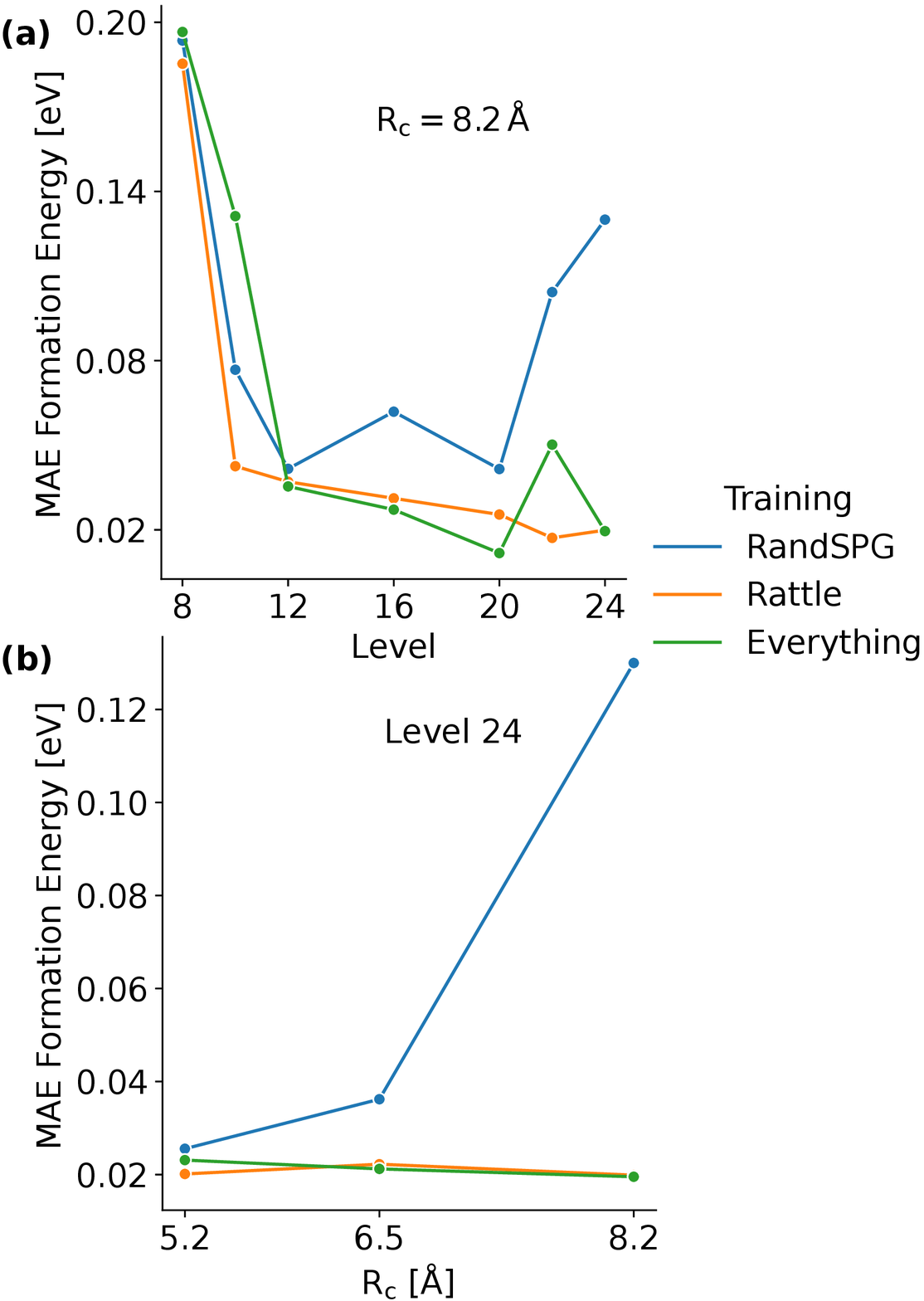}
    \caption{
        Mean absolute error (MAE) in the vacancy formation energy, $E^f_\protect\mathrm{vacancy}$, averaged over the four structure prototypes: hcp, fcc, dhcp and bcc.
        (a) MAE as function of potential level for potentials fitted with $R_c=\SI{8.2}{\AA}$ for the \spg~, \rattle~and \everything~sets.
        The best potential achieve errors as low as \SI{0.02}{eV} whereas the potential fitted \spg~shows increasing errors after level 20, a sign of overfitting due to its smaller size compared to \rattle~and \everything.
        (b) MAE as a function of $R_c$ for potentials of level 24 for the same training sets.
        The error increases on the smaller training set \spg{}, but is constant or decreases on the larger training sets \rattle{} and \everything{}.
    }
    \label{fig:vacancy_error}
\end{figure}

The mean absolute error (MAE) over the structure prototypes is shown in fig.\,\ref{fig:vacancy_error} as a function of MTP level and cutoff for the representative sets \spg{}, \rattle{} and \everything{}.
Section \ref{sec:appendix-vacancy} shows the error on all other sets as well.
The error decreases with level and saturates around \SI{0.02}{eV} before increasing again in case of the \spg{} set.
As this set is smaller than \everything{} and \rattle{}, it indicates overfitting on this set.

Shown in the bottom of fig.\,\ref{fig:vacancy_error} is the MAE as a function of cutoff at the highest fitted potential level.
The smaller set \spg{} shows increasing errors with cutoff and the (not shown here) minimized sets \volmin{} and \cellmin{} reproduce the same trend as the \spg{} set.
In contrast the larger sets \rattle{} and \everything{} show constant or decreasing errors with cutoff.
The other larger sets \hydro{} and \shear{} follow this trend.
We interpret this to mean that increasing the cutoff on smaller, less diverse training sets leads to the potentials only seeing more of the same environments, and hence overfitting, whereas it is more useful on larger, more diverse training sets, where potentials can pick up on more details that fall out of smaller cutoffs.

\begin{figure}
    \centering
    \includegraphics[width=\textwidth]{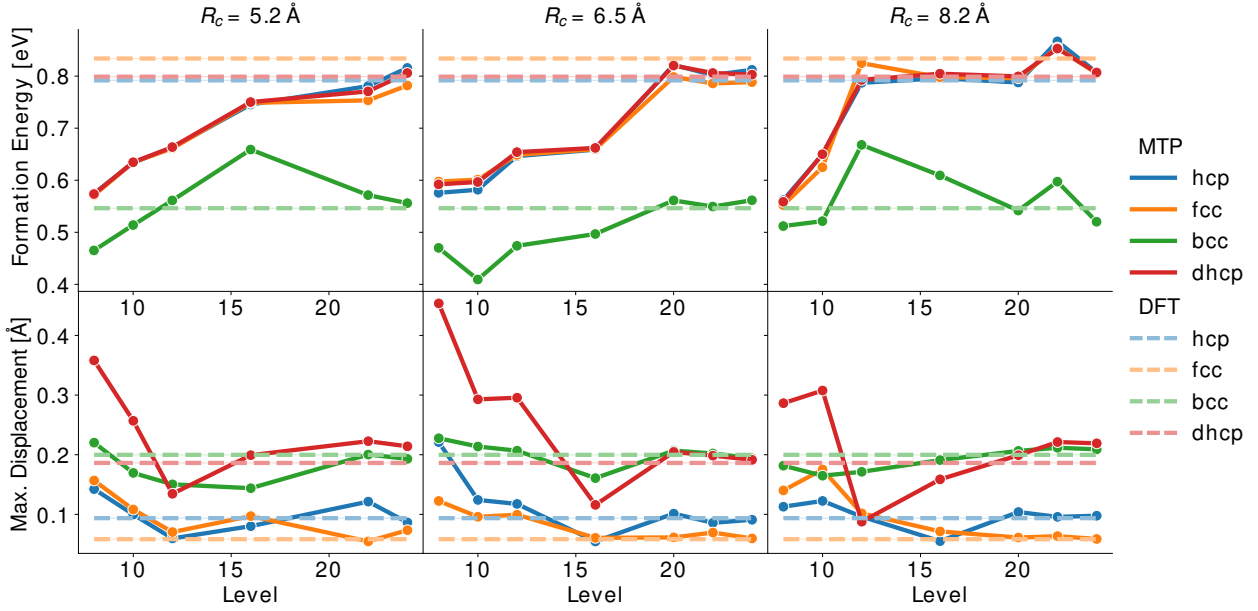}
    \caption{
        Comparisons between DFT reference and MTP prediction in formation energy and maximum displacement during minimization of vacancies.
        Here only potentials fitted to \protect\everything{} are shown,
        horizontal dashed lines are DFT references.
        (top row) Predicted energy of vacancy formation.
        (bottom row) Maximum displacement during minimization of the vacancy super cell across atoms and dimensions.
    }
    \label{fig:vacancy_max_disp_everything}
    \label{fig:vacancy_formation_everything}
\end{figure}

The absolute predictions of formation energy for just the potentials fit to the \everything~set are also shown in the top row of \cref{fig:vacancy_formation_everything} as a function of potential level.
Horizontal lines are the DFT reference energies.
Generally all three cut-off potentials manage to adequately (\SI{\sim 0.1}{eV}) describe the vacancy even at moderate potential levels ($\sim 16$), except for the bcc vacancy.
However, the potentials fit with $R_c = \SI{8.2}{\angstrom}$ manage good performance already at level 12. 
We therefore can conclude that at least potential cutoffs beyond \SI{5.2}{\angstrom} and potential levels beyond 16 are needed.

Next to the vacancy formation energies we also tested the performance of the potentials at predicting structural relaxations.  
The bottom row of \cref{fig:vacancy_max_disp_everything} shows the maximum displacement (across all atoms and dimensions) during the minimization of the vacancy super cell.
As in \cref{fig:vacancy_formation_everything} already moderate potential levels manage adequate agreement and, again, lower cutoffs generally reduce predictive power.  
This suggest that the potentials with high basis set and cutoff also correctly relax the structure around the defect.

\subsubsection{Planar Defects}\label{sec:gb}

We calculate surface energies for the HCP $(0001)$, $(10\bar 10)$ and $(1\bar 1 20)$ surfaces as well as a $\Sigma 13$, two configurations of $\Sigma 19$ (various rotations around $[0001]$)\,\cite{Wang1997}, $\Sigma 7b$ ($[12\bar30](001)$) grain boundaries and a basal reflection twin with lattice parameter $a=\SI{3.195}{\angstrom}$ and $c/a = 1.624$.
The $(10\bar 10)$ surface supports two different structures depending on which half plane terminates the surface and we have included both structures here.
The Mg $\Sigma 7b$ also has a second realization, called $T$-type\cite{Wang2012}, but they are very close in energy and they are not included in the calculations below, though we have verified similar accuracy on both structures independently.
All surface slabs are at least \SI{17}{\angstrom} thick, more than twice the largest potential cutoff.
The internal degrees of freedom are relaxed for each potential with the lateral lattice parameters fixed.
The two grain boundary structures are relaxed normal to plane in DFT first.
For each simulation we calculate the excess defect energy according to
\begin{equation}
    E_\mathrm{defect} = \frac{1}{A} \left(
    E_\mathrm{supercell} - \frac{N_\mathrm{supercell}}{N_\mathrm{reference}}
                            E_\mathrm{reference}
    \right),
\end{equation} 
where the reference is bulk hcp Mg with the above mentioned lattice parameters and $A$ is the cross section of the supercell.

\begin{figure}
    \centering
    \includegraphics[height=.7\textheight]{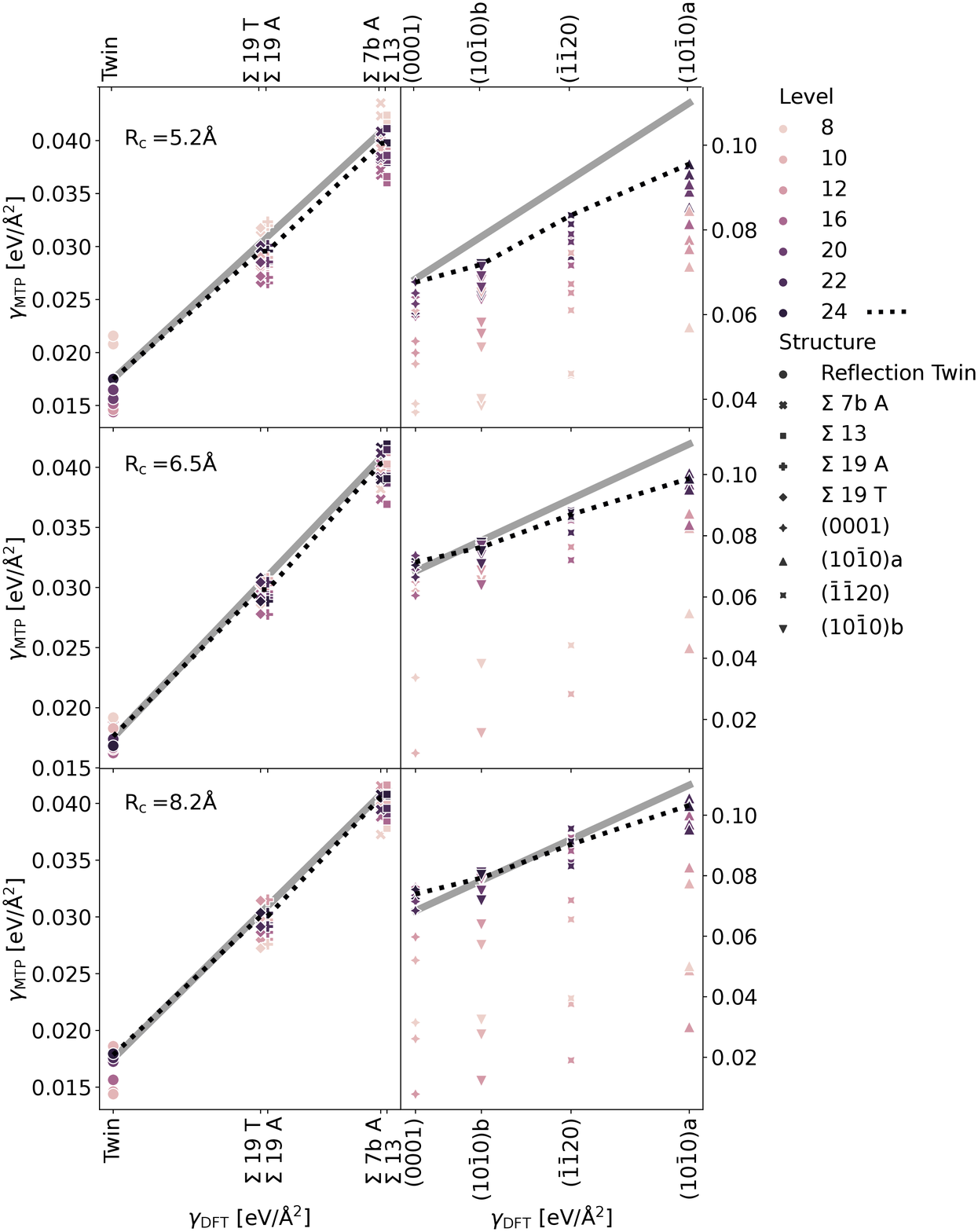}
    \caption{
        Correlation plot of planar defect excess energies; DFT vs. MTP predicted.
        Each point is a potential fit with the given cutoff and level highlighted by the hue.
        The grey line highlights perfect correlation;
        the dotted line connects the points of potentials with level 24.
        Compared to grain boundaries the initial accuracy of the potentials on surfaces is very poor, but increases substantially with higher levels.
        Note however, that except for the highest cutoff, even the highest levels systematically underestimate the surface energies. 
    }
    \label{fig:planar_defects_rmse}
\end{figure}

Differences between DFT and each of the potentials are plotted in \cref{fig:planar_defects_rmse} for the final potentials fitted to \everything.
All three cutoffs underestimate the surface energy at low potential levels, but show improvement with increasing level.
Interestingly lower cutoffs seem to systematically underestimate the surface energy, whereas higher cutoffs have larger errors at low level, but get much more precise than the low cutoff at larger levels.
The two grain boundaries are well described at all potential levels and cutoffs.
For the \everything{} set and $R_c = \SI{8.2}{\angstrom}$ the error on the surface structures is generally below \SI{100}{meV/\angstrom^2} and below \SI{10}{meV/\angstrom^2} for levels larger than 16.
The error on grain boundary structures is even an order of magnitude lower, except for the minimized training sets where the high-level potentials fail.

Additionally also generalized stacking fault energies and decohesion curves were tested with data provided by Stricker~\emph{et al.}\,\cite{Stricker2020Data} in section~\ref{sec:gsfe}.
We find very good agreement for all structures with errors below \SI{5}{meV/\AA^2} even for the lowest level potentials, except for the surface energies as already indicated in fig.\,\ref{fig:planar_defects_rmse}.

This section shows that our potential can faithfully reproduce planar defects and surfaces without having seen them during training explicitly.
We interpret this important fact as an indication that the \spg{} set is complete in the sense that it contains most of the local environments that are present in planar defects.

\subsection{Testing at high Temperature and Pressure}

As a final test of the stability of the fitted potentials over a wide range of thermodynamic conditions we calculate the thermal expansion and isothermal compressibility. 
Since hcp is the only thermodynamically stable configuration at zero or low pressures, results are shown here only for this structure. However, we also ran the same simulations for bcc, dhcp and fcc and have verified that MD simulations are always stable.
The simulations are run with $4\times4\times4$ unit cells for \SI{1e6} MD steps.

\begin{figure}
    \includegraphics[width=.8\textwidth]{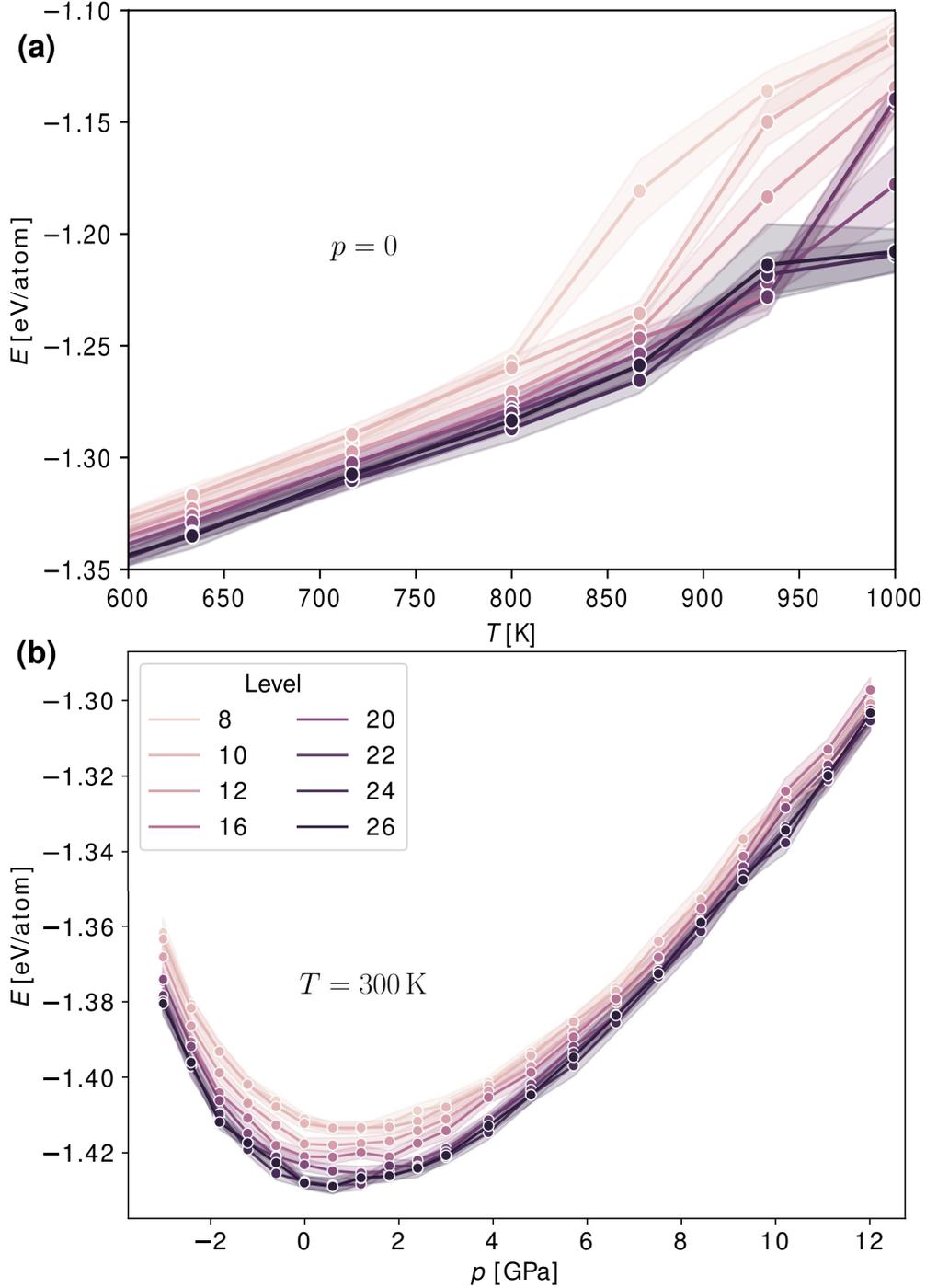}
    \caption{
        Results from $NPT$ MD simulations for potentials fit to \everything{} and $R_c = \SI{8.2}{\angstrom}$.
        Simulations for the other two cutoffs were run, but do not show significant differences from the ones shown. 
        (a) Average energy as a function of temperature at $p=\SI{0}{GPa}$.
            Melting is indicated by the abrupt change the temperature range \SIrange{800}{900}{K} depending on the potential level.
        (b) Average energy as a function of hydrostatic pressure at $T=\SI{300}{K}$.
    }\label{fig:properties}
\end{figure}

\Cref{fig:properties} shows the change of internal energy of hcp Mg with temperature at zero pressure and with pressure at $T=\SI{300}{K}$.
Simulation boxes remain stable until around \SIrange{800}{900}{K} depending on potential level, after which melting occurs.
While we did not carry out detailed calculations to precisely determine the melting point, this range is in good qualitative agreement with the experimental melting temperature of \SI{923}{K} reported for Mg, see e.\,g.\ \cite{nayebphase}.
\Cref{fig:properties} only shows potentials with a cutoff of \SI{8.2}{\angstrom}, but the cutoff appears to have no effect on the description of the simple thermodynamic state variables investigated here.
On the other hand the potential level seems to induce a shift in the finite temperature energy of a few \SI{10}{meV}.
We speculate that this is due to slightly larger forces constant errors at lower levels, which means slightly different heat capacities.

We investigate a pressure window from \SIrange{-3}{12}{GPa}.
Instabilities occur at large tensile pressures and elevated temperatures.
The lowest pressure that leads to unstable simulation boxes for hcp is \SI{-0.5}{GPa} at \SI{800}{K}, well above the tensile strength of pure magnesium.
We note that higher level potentials seem to be able to bear more tensile pressure than lower level ones before becoming unstable.

\subsection{Comparison of Potentials}\label{sec:stability_discusion}

\begin{figure}
    \centering
    \includegraphics[width=\textwidth]{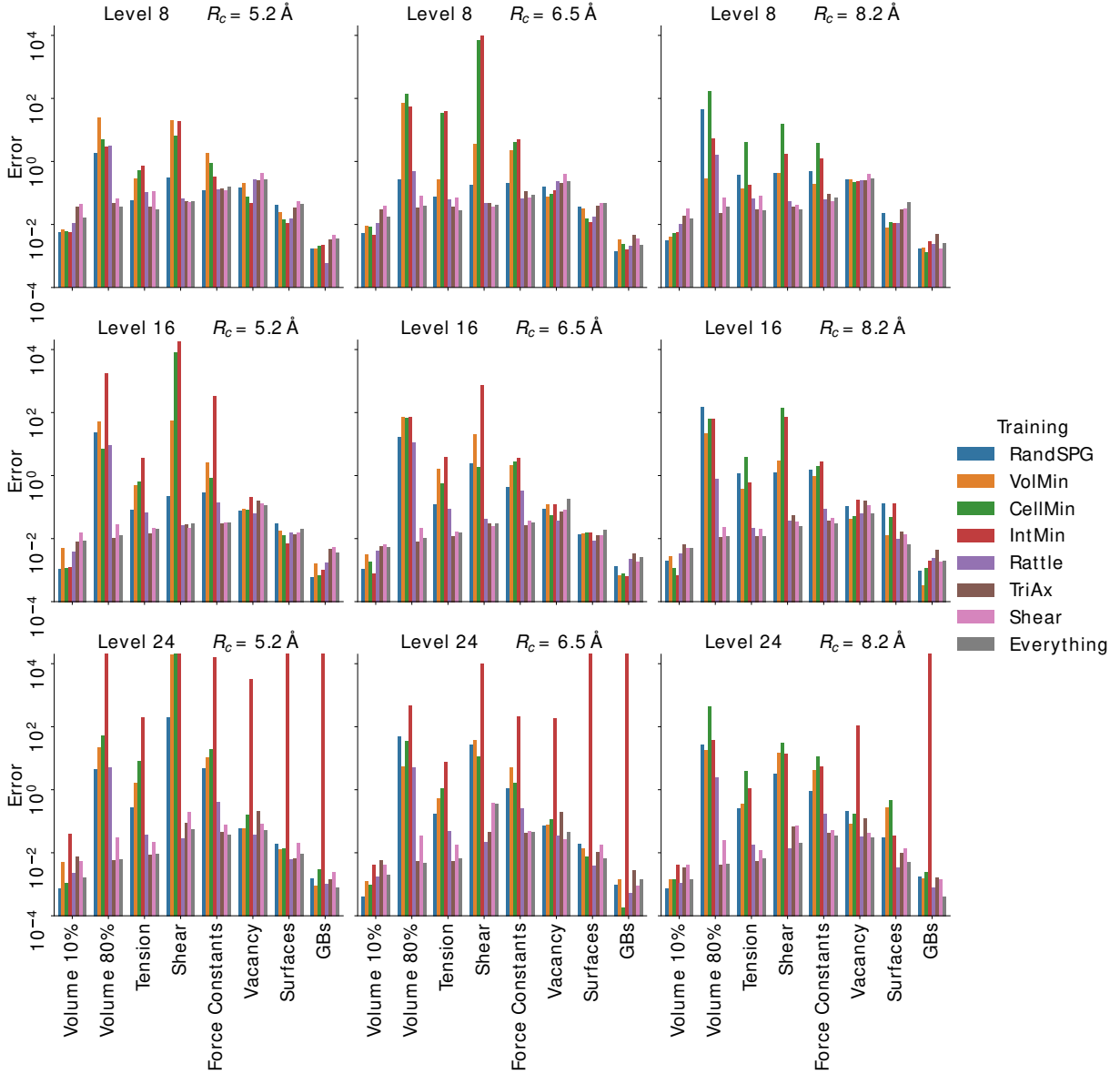}
    \caption{
        Errors on verification sets for three selected potential levels (8, 16, 24) and cutoffs (\SI{5.2}{\angstrom}, \SI{6.5}{\angstrom}, \SI{8.2}{\angstrom}).
        The x-axis enumerates the verification set, while the y-axis gives the logarithmic error on the set for the different training sets.
        For the sets "Volume", "Tension", and "Shear" the errors are the direct energy RMSE against DFT in eV;
        for "Force Constants" the errors the RMSE of the force constants in a super cell against DFT;
        and for the three defect sets "Vacancy", "Surface", "GBs" it is the RMSE of the respective energy excess of formation.
    }
    \label{fig:spider}
\end{figure}

We now collect the errors against DFT calculated in the previous sections for all training sets and test domains in \cref{fig:spider}.
Generally potentials fit to \spg~show average performance overall, except on tests with large strains. 
Potentials fit to the minimized sets (\volmin, \cellmin, \intmin) tend to perform badly, due to their reduced inherent dimensionality after the minimization steps.
While potentials fit to \everything~are not always the best performing potentials in each test case, they consistently rank among the top potentials.
For all defects at fixed potential level higher cutoffs give smaller errors, except for level 8, where it first increases for two of the minimized sets, \cellmin~and \intmin.
This aligns with our discussion of \cref{fig:energy_errors_sub_rmax}, that the low level descriptors are not flexible enough to make use of all the information available.
The minimized sets are shown to be overfit, giving completely nonsensical results on some of the verification sets, worsening with increasing level.
Comparing the two volume verification sets, we note that for the 10\,\% set the training sets follow the same trend in accuracy as in \cref{fig:energy_errors_sub_levels}, which we can understand since this is also the volume range present in each training set.
On the substantially larger 80\,\% set only \hydro~(as expected) and \shear~perform well.
It is not completely surprising, but interesting that the \shear~set outperforms the other sets so strongly, not having seen any (uniaxially) strained structures either, but the shear structures seem to carry at least some information also on strain.
In the single axis strained sets again, \hydro~and \shear~perform well, but also \rattle~is in between both sets.
Even though \rattle~only experiences strain and shear up to 5\,\% during training, it still seems to give adequate results on the verification up to 60\,\%.
On the other hand the \shear~set does not substantially improve the performance of the \everything~set anymore, but causes it to perform less well on different stackings in closed packed structures as discussed in Section~\ref{sec:volume} and \cref{fig:vol_range_rmse_comparison,fig:mtp_stacking_failure}.
In the final potentials provided in the supplementary we have therefore excluded it again.

Vacancies appear least well described of all defects checked, but this is also due to that fact the errors cited for the planar defects and surfaces are given normalized to the area.
Surfaces however are less well described than the grain boundaries, since they are included in the training set only indirectly due to the \volmin~set, as indicated earlier.
Still the potentials with highest level achieve less then \SI{10}{meV/angstrom^2} error on them.
For grain boundaries the accuracy is even in the order of \SI{1}{meV/angstrom^2}.
We speculate that accuracy on the surfaces could be improved by cutting the randomly generated bulk structures of the \spg~set without compromising the unbiased sampling.

\subsection{Comparison to Active Learning}\label{sec:al}

We want to further show that potentials fitted to a wide range of physically inspired structures are able to be reliably transferred to structures outside their original training domain. The active learning scheme as implemented by \mlip\cite{Podryabinkin2017} is reviewed in appendix~\ref{sec:appendix_al_review}.
We present calculations that show the potentials fitted in this work would remain unchanged under an active-learning scheme for applications investigated here.

We run MD for \SI{100}{ps} on four different structures under the active learning regime provided by \mlip. 
The structures are liquid Mg under high pressure ($T = \SI{2500}{K}$ and $p=\SI{5}{GPa}$, to keep the simulation box stable); an HCP Mg vacancy in a $4 \times 4 \times 4$ super cell at moderate temperatures and ambient pressure ($T$ in 100, 400, 700K); an HCP twin boundary and a $\Sigma 7\mathrm{b}$ grain boundary at \SI{600}{K} and ambient pressure; and finally fcc and dhcp structures at ambient temperature and pressure.
The super cells are adjusted for each potential to allow for at least twice the potential cutoff radius to fall within the periodic boundaries.
We set selection, $\gamma_\mathrm{select}=1.001$, and extrapolation thresholds, $\gamma_\mathrm{break}=5$, to catch any extrapolation.
They define when a structure is selected for active learning or when the simulation is deemed to inaccurate, c.\,f.\ appendix~\ref{sec:appendix_al_review} for their precise definitions.
Within the given time frame \emph{none} of the potentials trained on the \everything~set encountered structures where $\gamma > \gamma_\mathrm{select}$, i.e. there was no evidence of extrapolation.
Appendix~\ref{sec:appendix_al_review} also shows exemplary calculations of the extrapolation grade over simulation time for potentials fitted on subset of \everything.

We can now verify our hypothesis explaining the success of the potential in Section~\ref{sec:gb}; if the accuracy were due to good extrapolation, the active learning would still flag the structures as outside of the active set.
Since that did not happen, there must be local environments in the training set that are sufficiently close in descriptor space to allow interpolation of the defect structures.

With respect to the failure of low level potentials to differentiate closed packed structures as discussed in Section~\ref{sec:volume}, we can now clearly state that this is an interpolation error, since dhcp and fcc structures were not identified as extrapolation by \mlip.
This is in contrast to Zeni \emph{et al.}~\cite{Zeni2022}.
There, they fit Atomic Cluster Expansion (ACE)\cite{Drautz2019} potentials to a variety of systems and e.\,g.\ show that potentials fit to liquid water often enter the extrapolative regime when applied to equilibrium ice structures.
Since ACE and MTP potentials both completely span the space of local atomic environments\,\cite[Appendix B]{Dusson2022}, we believe our results and approach are also transferable to ACE potentials.
That our potentials do not extrapolate on the structures tested above, whereas they do in the study of Zeni \emph{et al.}\,\cite{Zeni2022}, shows that the choice of the structure set is critical.

On the other hand the original question they set out to answer and that has been discussed in the literature before is how to tell whether potentials are reliable or not in a given application.
The distinction of extrapolation vs.\ interpolation is then only a proxy for this.
We do not provide an answer, but our results indicate that this question is not settled yet.

From this discussion we are confident that the construction of our training set is complete and the potentials can be directly used in applications with respect to the tests performed here.
Still, users that wish to use the potentials on defects we have not verified, i.\,e.\ on isolated dislocation or cracks may need to do additional testing.
As for dislocations, given that the $\Sigma 7b$ interface is comprised of dislocation cores\,\cite{Carter2014} we do not expect significant failure.
In any case we expect our structure sets to be excellent starting points for additional active learning where necessary.

\subsection{Transferability}

For the purpose of this work take a \emph{transferable potential} to be one that describes not only structures it has been fitted to, but also any other structures it might be applied to or is at least well-behaved.
This is true for most empirical potentials, but not for most machine learning potentials, and is often cited as their strength.

It is naturally not possible to exhaustively test this property on any given potential, though the preceding sections show that the potentials fitted here describe defects very well without being fitted on them.
On the other hand, it is at least necessary that a transferable potential can adequately describe the structures of the training set of another potential.
We have tested this condition here by applying two other machine learning potentials for Mg (a HDNNP by Stricker \emph{et al.}\,\cite{Stricker2020} and a recent RANN by Barret \emph{et al.}\,\cite{Barret2022}) on the training set presented here and vice-versa and find that our potentials achieve similar RMSE on other training sets as on their own sets.

The network based potentials however perform much worse on the broad \everything{} training set than on their own respective sets.
This finding should not be mistaken to mean those potentials or their underlying formalism perform badly or are ill-suited to the study of Mg and its defects.
In fact they perform very well on the defects they were used to study.\,\cite{Barret2022,Liu2023}
It merely indicates that they are not (and were not meant to be) general potentials or at least do not transfer as well as potentials fitted to more extensive training sets as the one presented here.
The numerical results of this comparison are summarized in the Appendix~\ref{sec:app_transfer}.

\section{Conclusion}\label{sec:conclusion}

We demonstrate that an unbiased, systematic construction of the training set, which covers \emph{all} bulk crystal symmetries instead of just low energy structures, allows the successful construction of ML-potentials that have a high degree of transferability, in particular to bulk defects.
Importantly, the training database does not include any explicit defect structures or additional data from active learning, see \cref{fig:planar_defects_rmse,fig:vacancy_formation_everything}.

In \cref{fig:energy_errors_sub_rmax,fig:planar_defects_rmse} we show that the determination of the cutoff radius warrants more care than is sometimes paid in literature: A large enough cutoff radius is crucial for the transferability of the potential to structures not included in training (in this case surfaces).
To fully utilize this benefit potentials of higher level (basis set) are required.
In the future it may hence be worthwhile to separately increase the number of radial functions keeping the maximum tensor power in the basis functions (i.\,e.\ the body order) constant to inspect and utilize this effect.

In Section~\ref{sec:al} we further show that active learning does not provide additional benefits once a sufficiently diverse training set is considered.
Additionally, we show that machine learning potentials can give non-optimal results even when they are not extrapolating.
This is an important statement as it is often implicitly or explicitly assumed by researchers developing machine learning and active learning formalisms that interpolation errors can be neglected.
The question how to treat them hence continues to be an open question.

We expect our observations hold for general MLIPs with descriptors that form a complete basis, but at least for the ACE, since MTP descriptors can be expressed in the ACE basis as well.

In practical terms we provide a general purpose potential for Mg, that describes the equilibrium hcp and high-pressure bcc phases, and also planar and point defects.
We recommend levels higher than 16 with $R_c = \SI{8.2}{\angstrom}$, to avoid wrong (zero) stacking energies predicted for low level potentials\footnote{See \cref{fig:mtp_stacking_failure}.}.
Lower levels and cutoffs may be used to save computational resources, though care has to be taken so that the interpolation errors do not jeopardize the application.
For these cases it may be more advisable to custom fit lower level potentials to narrower training sets.

The approach to construct unbiased and physical data sets by systematically sampling bulk structures over all bulk symmetries can be straightforwardly applied to other materials systems.
The construction approach outlined in this study facilitate the construction of such data sets in a systematic and automated fashion.
We have shown that potentials fitted to training sets of this kind transfer better to training sets of other potentials than vice-versa.
This opens the route towards a largely automatized generation of general, transferable and accurate potentials. 

\section*{Data Availability}

The full DFT training set and resulting MTP potential files for the potentials fitted to \everything{} with and without the \shear{} set can be accessed at \url{https://doi.org/10.17617/3.A3MB7Z}.
Under\\
\url{https://github.com/eisenforschung/magnesium-mtp-training-data}
we have also added example jupyter notebooks and a \pyiron~project that shows how to access the data, generate similar training sets and run simple simulations with \lammps.

\begin{acknowledgments}

The authors acknowledge funding by the Deutsche Forschungsgemeinschaft (DFG, German Research Foundation) through the Collaborative Research Center 1394 (SFB 1394, No. 409476157) and Project No. 405621160.
EB further acknowledges support from the European Research Council (ERC) under the European Union’s Horizon 2020 research and innovation programme (grant agreement No. 725483).
We also thank Ralf Drautz, whose potential fitting experience benefited us during this study, Prince Matthews for providing grain boundary structures and Markus Stricker for help with setting up their neural network potential and providing defect structures.

\end{acknowledgments}

\bibliography{bib}

\appendix

\section{Cost-Accuracy Tradeoff}\label{sec:appendix-pareto}

\begin{figure}
    \centering
    \includegraphics[width=\textwidth]{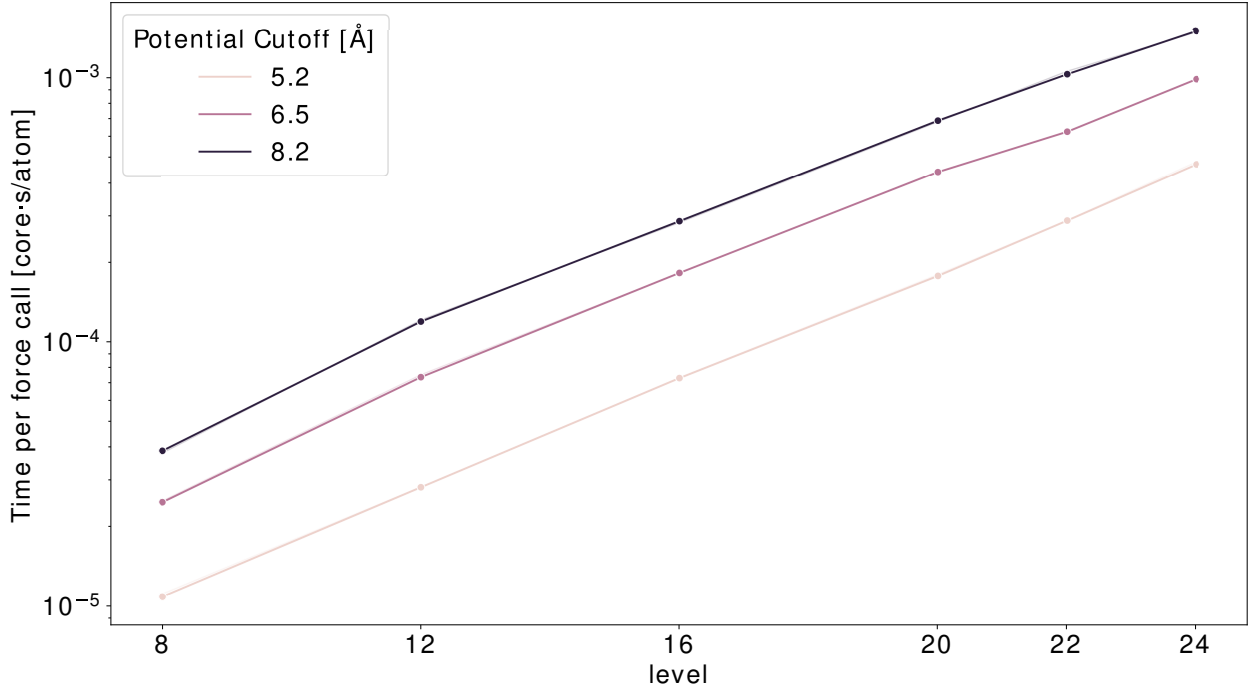}
    \caption{Time per force per atom versus potential level for different cutoffs.  Note the log-linear trend due to the exponential number of parameters as a function of level.}
    \label{fig:runtime}
\end{figure}

\Cref{fig:runtime} shows the cost per force call per atom as a function of potential level and cutoff radius.  
There is a log-linear trend in the runtime versus level as expected, since the number of free parameters in the potential increases exponentially with the level.  
Increments in radial cutoff only change the prefactor of the scaling, which is also expected of local models.  

\section{Errors on Vacancy Formation Energy for all Training Sets}\label{sec:appendix-vacancy}

\begin{figure}
    \centering
    \includegraphics[height=.2\textheight]{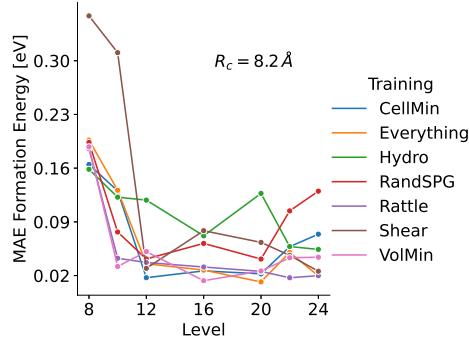}
    \caption{
    Mean absolute error (MAE) in the vacancy formation energy, $E^f_\protect\mathrm{vacancy}$, averaged over the four structure prototypes: hcp, fcc, dhcp and bcc,
    as function of potential level for potentials fitted with $R_c=\SI{8.2}{\AA}$ for all sets except \intmin{}.
    }
    \label{fig:vac_level_all}
\end{figure}

\begin{figure}
    \centering
    \includegraphics[height=.2\textheight]{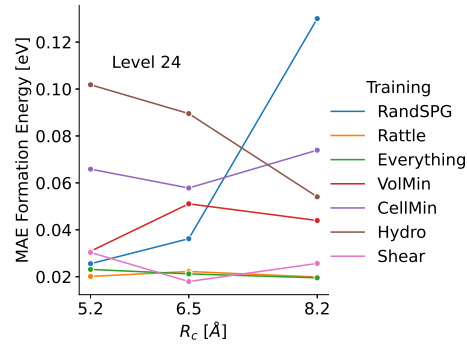}
    \caption{
    Mean absolute error (MAE) in the vacancy formation energy, $E^f_\protect\mathrm{vacancy}$, averaged over the four structure prototypes: hcp, fcc, dhcp and bcc, as a function of $R_c$ for potentials of level 24 to all training sets, except \intmin.
    }
    \label{fig:vac_rmax_all}
\end{figure}

\begin{figure}
    \centering
    \includegraphics[height=.2\textheight]{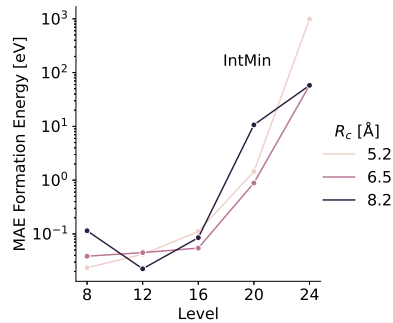}
    \caption{
    Mean absolute error (MAE) in the vacancy formation energy, $E^f_\protect\mathrm{vacancy}$, averaged over the four structure prototypes: hcp, fcc, dhcp and bcc, for the potentials fitted to \intmin{}.
    At high levels the potentials clearly fail, indicated that the training set is not diverse enough for the model complexity.
    }
    \label{fig:vac_intmin}
\end{figure}

In figures \ref{fig:vac_level_all}, \ref{fig:vac_rmax_all} and \ref{fig:vac_intmin} the full data behind figure \ref{fig:vacancy_error} is shown.
The minimized sets \volmin{}, \cellmin{} and \intmin{} follow the same trend as \spg{}, whereas \hydro{}, \shear{} follow the trend of \rattle{}.
The \intmin{} set is clearly shown to be insufficient for potentials with high levels.

\section{Stacking Faults and Decohesion Curves}\label{sec:gsfe}

\begin{figure}
    \centering
    \includegraphics[width=\textwidth]{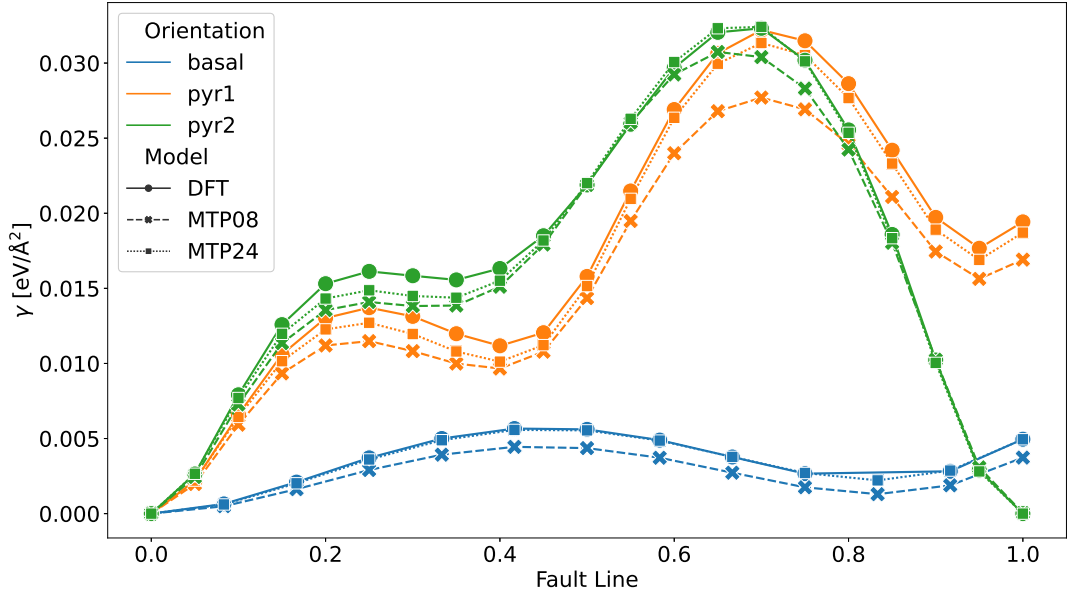}
    \caption{
        GSFE curves for the basal, pyramidal 1 and pyramidal 2 stacking faults in hcp Mg calculated with DFT (solid lines) and two potentials with $R_c=\SI{8.2}{\AA}$ and level 8 and 24 (dashed and dotted lines).
        All three description are qualitatively the same with the level 24 slightly better than level 8.
        The errors for all structures compared to DFT are below \SI{5}{meV/\AA^2}.
    }
    \label{fig:gsfe}
\end{figure}

\begin{figure}
    \centering
    \includegraphics[width=\textwidth]{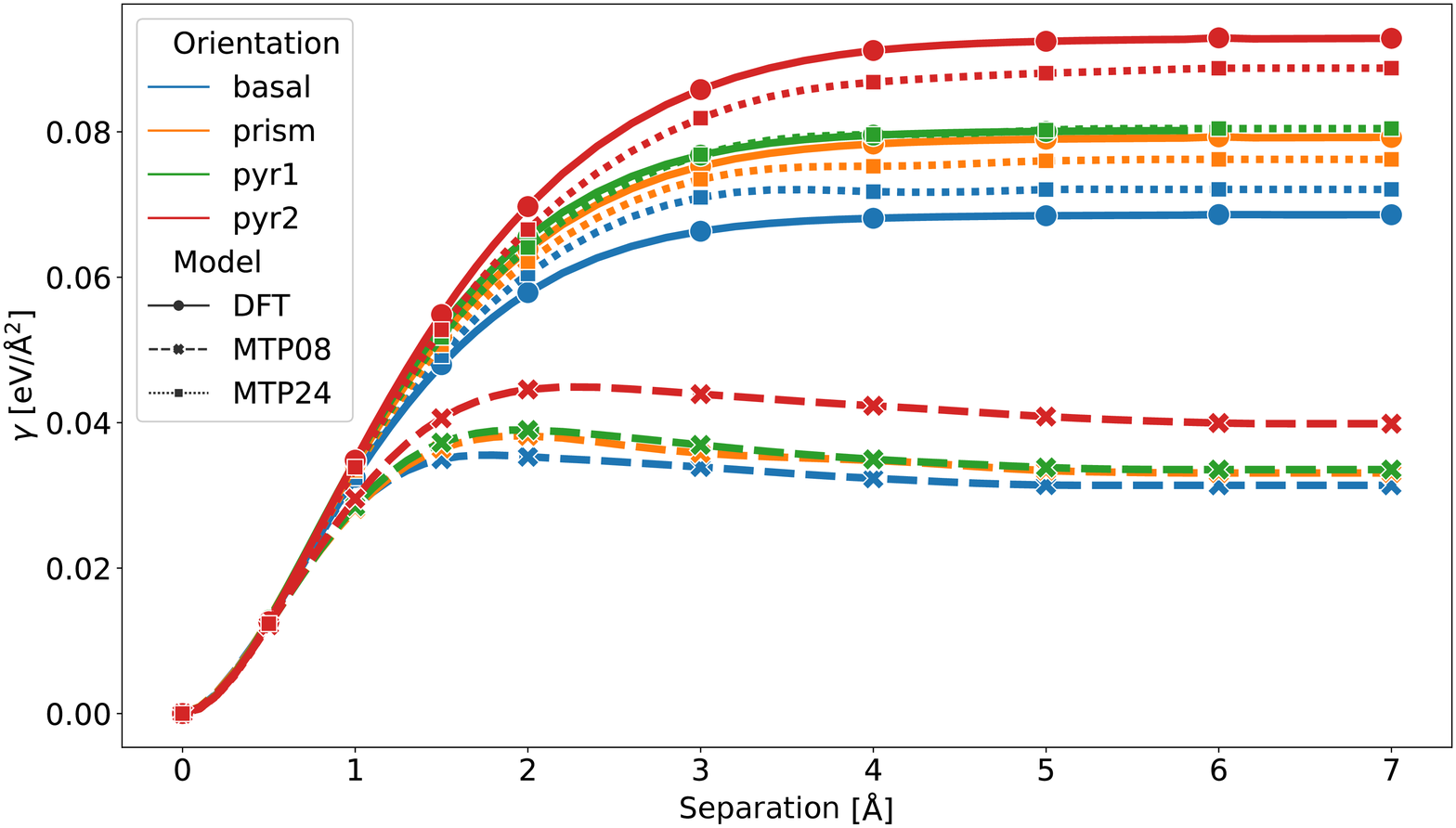}
    \caption{
        Decohesion curves for the basal, pyramidal 1, 2 and prismatic surface in hcp Mg calculated with DFT (solid lines) and two potentials with $R_c=\SI{8.2}{\AA}$ and level 8 and 24 (dashed and dotted lines).
        Markers are placed only as an aid to the eye, data points are denser.
        At short ranges both potentials give a correct description, but the level 8 potential (dashed lines with crosses) fails quickly due its severe underestimation of the surface energies.
    }
    \label{fig:decohesion}
\end{figure}

Based on the training data from Stricker~\emph{et al.}\,\cite{Stricker2020Data} we have compared also the generalized stacking fault energy (GSFE) and decohesion curves along various orientations in hcp. We picked two potentials fitted with $R_c=\SI{8.2}{\AA}$ and level 8 and 24 as representative for the potentials fitted here.
Figure\,\ref{fig:gsfe} shows the 
The results are summarized in figs.\,\ref{fig:gsfe} and \ref{fig:decohesion}.
They show the energy difference along the stacking fault path or decohesion.
Since DFT and the potentials predict slightly different cohesive energies for the bulk structures, we have subtracted this difference as only the energy barriers along the path are of physical significance.
For the level 24 potential all structures are in the range of a few \SI{}{meV/\AA} 
In contrast the level 8 potential does not give an adequate description of the decohesion.
This is because they underestimate the surface energies as already noted in fig.\,\ref{fig:planar_defects_rmse}.
On the other hand they also do well on the GSFE curves.
For technical reasons\footnote{
    With \vasp{} we experienced crashes using the (rather high) $k$-point sampling that we employed in this work.
    \vasp{} and \sphinx{} use a different energy reference, but that is not relevant here, since we are calculating energy differences only.
} the DFT calculation for comparisons were carried out with the \sphinx~DFT code, but using the \vasp~provided pseudopotential files and equivalent settings as the rest of this paper.

\section{Transferability Comparison}\label{sec:app_transfer}

Table~\ref{tab:transfer} compiles the results of the transferability tests between a MTP of level 24 with $R_c = \SI{8.2}{\AA}$ and the HDNNP\,\cite{Stricker2020} and RANN\,\cite{Barret2022} network potentials mentioned in the main text.
For the calculation of the RMSE we have used the structures and corresponding DFT energies of each publication.

Since the HDNNP training set also used \vasp{}, we have checked that our convergence settings and theirs produce matching DFT energies on a sub set of their training set.
For the RANN training set we shifted the DFT energies to agree with the experimental cohesive energies of Mg, as done in their parametrization of the potential.

For the full \everything{} set the two network potentials fail, so we also checked a subset with only those structures that have negative energies to avoid those with atoms very close to each other.
For the HDNNP training set, we found the errors of all potentials to be dominated by the dimer curve and the long distance part of the $E$--$V$ curve, so we also checked subsets without those structures, and similarly also for the RANN training.

Errors on structures with very large separation between atoms are also affected by the fact that all three potentials give slightly different energies of the isolated atom (as they are fit to different references).
The failure of the RANN potential on the set with the dimer structures is only due to some (unlikely to be relevant) structures with a Mg--Mg separation of less than \SI{1}{\AA}.
The largest contribution to the RMSE of the potential fitted here is in surface structures, which could be remedied in future work by systematically including surfaces, as sketched in Section~\ref{sec:gb}.

\begin{table}[h]
    \centering
    \caption{
        RMSE in m\,eV/atom for three potentials mentioned in the main text tested on each training set.
        Asterisks (*) indicate errors in excess of \SI{1}{eV/atom}.  
    }
    \begin{tabular}{c|rr|rrr|rr}
        \toprule
            &  \multicolumn{2}{c|}{\everything} & \multicolumn{3}{c|}{Stricker \emph{et al.}\,\cite{Stricker2020Data}}  & \multicolumn{2}{c}{Nitol \emph{et al.}\,\cite{Barret2022}} \\
            & all & $E<0$                       & all & no dimer & no dimer                                             & all & no dimer \\
            &     &                             &     &          & no $E$--$V$                                          &     &          \\
        \midrule
         MTP24  &   4.8 &   4.2 &  87.2 &   9.1 &   2.7 &  21.7 &   7.0   \\
         HDNNP  &     * & 400.2 & 157.9 &  19.3 &   1.3 &  71.1 &  72.1   \\
         RANN   &     * & 387.1 &     * &  37.1 &   3.9 &   0.4 &   0.4    \\
         \bottomrule
    \end{tabular}
    \label{tab:transfer}
\end{table}

\section{Active Learning}\label{sec:app_al}

\subsection{Active Learning Scheme}\label{sec:appendix_al_review}

\begin{figure}
    \centering
    \def\svgwidth{.7\linewidth}
\begingroup%
  \makeatletter%
  \providecommand\color[2][]{%
    \errmessage{(Inkscape) Color is used for the text in Inkscape, but the package 'color.sty' is not loaded}%
    \renewcommand\color[2][]{}%
  }%
  \providecommand\transparent[1]{%
    \errmessage{(Inkscape) Transparency is used (non-zero) for the text in Inkscape, but the package 'transparent.sty' is not loaded}%
    \renewcommand\transparent[1]{}%
  }%
  \providecommand\rotatebox[2]{#2}%
  \newcommand*\fsize{\dimexpr\f@size pt\relax}%
  \newcommand*\lineheight[1]{\fontsize{\fsize}{#1\fsize}\selectfont}%
  \ifx\svgwidth\undefined%
    \setlength{\unitlength}{595.27559055bp}%
    \ifx\svgscale\undefined%
      \relax%
    \else%
      \setlength{\unitlength}{\unitlength * \real{\svgscale}}%
    \fi%
  \else%
    \setlength{\unitlength}{\svgwidth}%
  \fi%
  \global\let\svgwidth\undefined%
  \global\let\svgscale\undefined%
  \makeatother%
  \begin{picture}(1,1.41428571)%
    \lineheight{1}%
    \setlength\tabcolsep{0pt}%
    \put(0,0){\includegraphics[width=\unitlength]{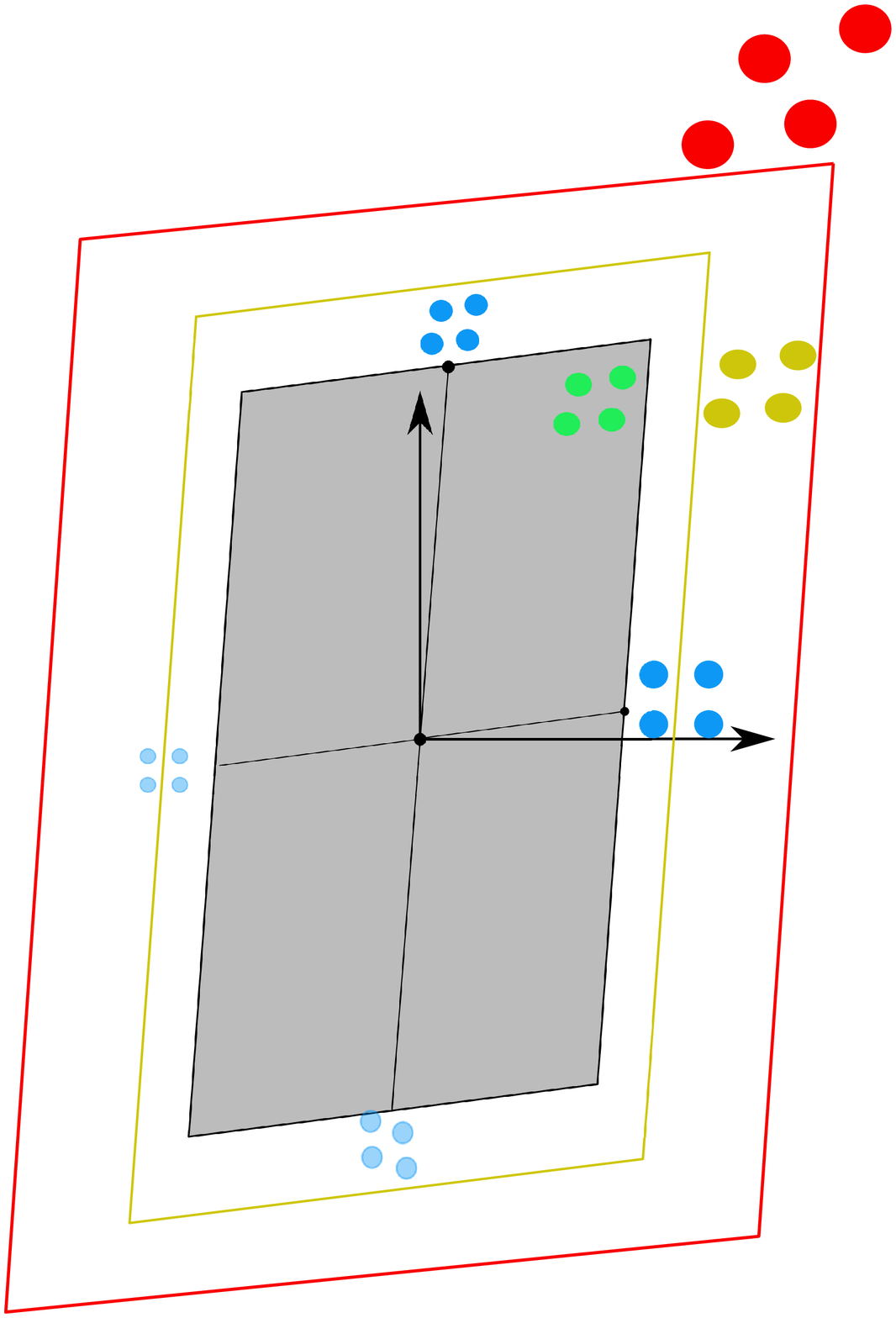}}%
    \put(0.5114761,0.58052063){\color[rgb]{0,0,0}\makebox(0,0)[lt]{\lineheight{1.25}\smash{\begin{tabular}[t]{l}Descriptor 1\end{tabular}}}}%
    \put(0.43614004,0.68912906){\color[rgb]{0,0,0}\rotatebox{90}{\makebox(0,0)[lt]{\lineheight{1.25}\smash{\begin{tabular}[t]{l}Descriptor 2\end{tabular}}}}}%
    \put(0.26125681,1.03933512){\color[rgb]{0,0,0}\makebox(0,0)[lt]{\lineheight{1.25}\smash{\begin{tabular}[t]{l}$\gamma_\mathrm{select}$\end{tabular}}}}%
    \put(0.21902755,1.11478293){\color[rgb]{0,0,0}\makebox(0,0)[lt]{\lineheight{1.25}\smash{\begin{tabular}[t]{l}$\gamma_\mathrm{break}$\end{tabular}}}}%
  \end{picture}%
\endgroup%

    \caption{
        Schematic illustration of phase space seen during training and active learning.
        Suppose our training set (blue structure) is made up of strained and sheared cubic structures. The phase space spanned by the active set then corresponds to the grey area.
        Structures inside the yellow borders (green) are assumed to be approximated well by the potential; between the yellow and the red border (yellow) are selected for further training and outside the red border (red) cause simulations to terminate.
    }
    \label{fig:al_carton}
\end{figure}

Suppose we have the list of $m$ basis functions calculated for a given atomic neighborhood $\left\{ b_i \right\}$ and we wish to know whether this configuration can be safely approximated by the potential or not.  
Novikov \emph{et al.}\,\cite{Novikov2021} answer this by defining an active set of configurations.
We can think of this set being chosen from all training structures such that they cover the widest range of phase space seen during training.\footnote{
    For detailed explanation of this algorithm see \cite{Novikov2021,Goreinov2010}
}
\Cref{fig:al_carton} shows a schematic illustration of the active set in solid blue, the covered phase space corresponds to the gray shaded area.
They then define the active learning state $A^{-1}$ as the matrix that projects our calculated coefficients $\left\{ b_i \right\}$ into the space spanned by the active set, i.e. we can obtain the basis coefficients of the given atomic neighborhood by matrix multiplication
\begin{equation}
    c = A^{-1} b.
\end{equation}
Now if all $\left\{ c_i \right\}$ are smaller than unity, \cite{Novikov2021} define this configuration to be within the interpolative region where small fitting errors can be assumed.  
If however one the of coefficients is larger than unity then the configuration is outside of the phase space region sampled during training and we might like to add it to the training set to improve the potential.
They quantify this notion by introducing
\begin{equation}
    \gamma(\mathrm{cfg}) 
    = \max_i \left| c_i \right|
    = \max_i \left| A^{-1}_{ij} b_j(\mathrm{cfg}) \right|
\end{equation}
where $\gamma \leq 0$ in the first case discussed above and $\gamma > 1$ in the second case.
The active learning regime implemented in \mlip then consists of running any desired simulation protocol but keeping tracking of $\gamma$ for all neighborhoods in the considered structure.
They define two thresholds, which we draw also in \cref{fig:al_carton} with yellow and red lines.
Exceeding the first (yellow), $\gamma_\mathrm{select}$, causes the whole structure to be written out for later consideration.
Exceeding the second (red), $\gamma_\mathrm{break}$, causes the whole simulation to be aborted.
When the simulation aborts due the latter case, the structures written in the first case can be used to enrich the training set and then re-run the simulation.
This is repeated until the full simulation runs without exceeding $\gamma_\mathrm{break}$.

\subsection{Example Simulation for Extrapolation Grade}

To illustrate the concept of the extrapolation grade we run three simulations of liquid Mg in an $NPT$ ensemble at \SI{2500}{K} and {5}{GPa} for \SI{1}{ps} and track the extrapolation grade over time.
We use $\gamma_\mathrm{select} = 1.001$ and $\gamma_\mathrm{break}=5$ as in the main text.
The potentials used for this illustration were fitted to a random sample of 10\,\% of \everything~with level 16 and $R_c=\SI{8.2}{\angstrom}$ to simulate an insufficient training set.

\begin{figure}
    \centering
    \includegraphics[width=\textwidth]{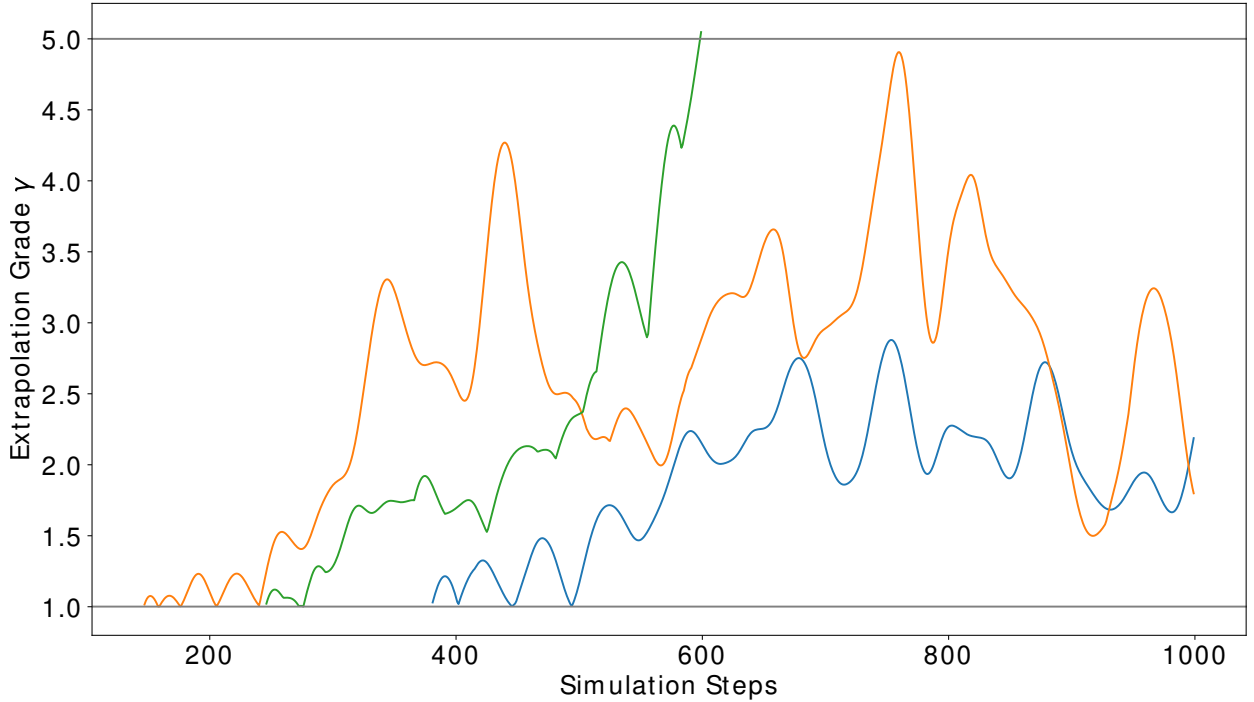}
    \caption{
        Extrapolation grade as a function of time during three $NPT$
        simulations of liquid Mg at \SI{2500}{K} and \SI{5}{GPa} using potentials fitted to insufficient training sets.
        Potentials can be seen interpolating initially, then exceed the selection threshold and even the breaking threshold in one case.
    }
    \label{fig:mv_grade_over_time}
\end{figure}

\section{Phonon Results for bcc Mg}\label{sec:app_phonon}

\begin{figure}
    \centering
    \includegraphics[width=\textwidth]{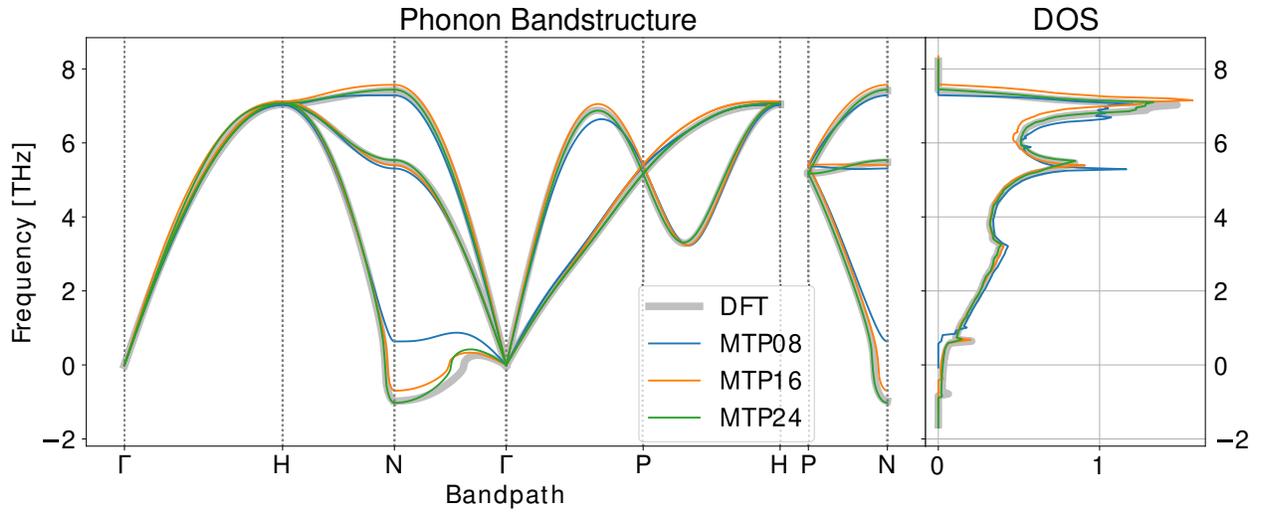}
    \caption{
        Phonon band structure and density of states of bcc Mg ($\Omega = \SI{22.83}{\angstrom^3/atom}$) calculated with MTPs fitted to \everything~with $R_c=\SI{8.2}{\angstrom}$ at levels 8, 16 and 24.
        Colored lines indicate the results from MTPs;
        the thick gray line indicates results from DFT.
    }
    \label{fig:bcc_phonon_min}
\end{figure}

\begin{figure}
    \centering
    \includegraphics[width=\textwidth]{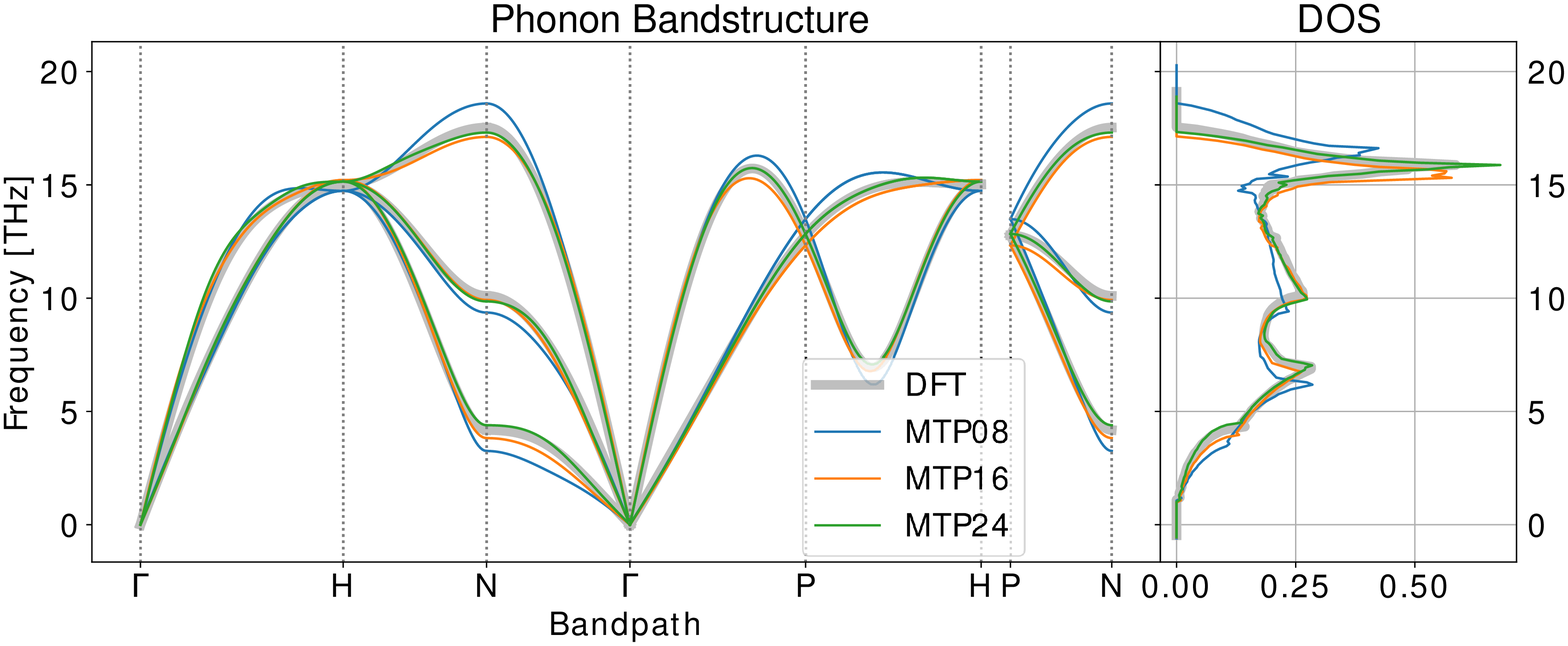}
    \caption{
        Phonon band structure and density of states of compressed bcc Mg ($\Omega = \SI{12}{\angstrom^3/atom}$) calculated with MTPs fitted to \everything~with $R_c=\SI{8.2}{\angstrom}$ at levels 8, 16 and 24.
        Colored lines indicate the results from MTPs;
        the thick gray line indicates results from DFT.
    }
    \label{fig:bcc_phonon_stable}
\end{figure}

\Cref{fig:bcc_phonon_min,fig:bcc_phonon_stable} show the phonon band structure of bcc Mg at $\Omega = \SI{22.83}{\angstrom^3/atom}$ and $\Omega = \SI{12}{\angstrom^3/atom}$ respectively.
It can be seen that also the phonons of the bcc polymorph are well described, even at volumes far from the equilibrium volume $\Omega_0$.
Note that in contrast to the potentials reviewed by Troncoso \emph{et al.}\,\cite{Troncoso2022} the potentials shown here correctly describe all the essential features as well as the negative part of the phonon spectrum in case of the minimized bcc structure.
The only exception is the level 8 potential that does not capture the dynamical instability of bcc Mg at ambient pressure.

\end{document}